\documentclass[11pt]{article}
\usepackage{verbatim,amsmath,amssymb}
\usepackage{epsfig,float,color}
\usepackage{epstopdf}
\usepackage{geometry}
\usepackage{setspace}
\usepackage{wrapfig}
\usepackage{hyperref}
\usepackage[utf8]{inputenc}
\usepackage{graphicx}
\usepackage{flushend}
\usepackage{tikz,pgfplots}
\usepackage[linesnumbered,ruled]{algorithm2e}

\usepackage[font={footnotesize}]{caption}
\usepackage[font={footnotesize}]{subcaption}
\usepackage{footnote}
\usepackage{multicol}
\usepackage{multirow}
\usepackage{enumitem}   
\usepackage{soul}
\usepackage{framed}
\usepackage[titletoc,title]{appendix}
\usepackage{bm}
\usepackage{siunitx}
\usepackage[makeroom]{cancel} 
\usepackage{natbib}
\usepackage{hyperref}
\usepackage{bigints} 

\newenvironment{rcases}
{\left.\begin{aligned}}
	{\end{aligned}\right\rbrace}

\definecolor{darkblue}{rgb}{0,0,1}
\hypersetup{pdftex=true, colorlinks=true, breaklinks=true, linkcolor=blue, menucolor=blue, citecolor=black, urlcolor=blue}

\newcommand{\trr}[1]{{#1}^{\!\top}}

\usepackage{tikz}
\usetikzlibrary{positioning, arrows.meta}
\tikzset{%
	myarrow/.style = {-Stealth, shorten >=5pt}
}

\usepackage{listings}
\definecolor{LightCyan}{rgb}{0.88,1,1}
\definecolor{mygreen}{RGB}{28,172,0} 
\definecolor{mylilas}{RGB}{170,55,241}
\newcommand*{\DrawBoundingBox}[1][]{%
	\draw [red, very thick, #1]
	([shift={(-1pt,0pt)}]current bounding box.north east)
	rectangle
	([shift={(1pt,1.5pt)}]current bounding box.south west);
}

\begin{document}
	
	\begin{center}
		\Large{\bf{$\mathtt{TOPress}$: a MATLAB implementation for topology optimization ofstructures subjected to design-dependent pressure loads}}\\
		
	\end{center}
	
	\begin{center}

		\large{Prabhat  Kumar\footnote{pkumar@mae.iith.ac.in; prabhatkumar.rns@gmail.com}}
		\vspace{4mm}
		
		\small{\textit{Department of Mechanical and Aerospace Engineering, Indian Institute of Technology Hyderabad, 502285, India}}
			\vspace{4mm}
			
 Published\footnote{This pdf is the personal version of an article whose final publication is available at \href{https://link.springer.com/article/10.1007/s00158-023-03533-9}{Structural and Multidisciplinary Optimization}}\,\,\,in \textit{Structural and Multidisciplinary Optimization}, 
			\href{https://link.springer.com/article/10.1007/s00158-023-03533-9}{DOI:10.1007/s00158-023-03533-9} \\
			Submitted on 13~February 2023, Revised on 13~February 2023, Accepted on 23~February 2023

	\end{center}
	
	\vspace{1mm}
	\rule{\linewidth}{.15mm}
	{\bf Abstract:}
	  In a topology optimization setting, design-dependent fluidic pressure loads pose several challenges as their direction, magnitude, and location alter with topology evolution. This paper offers a compact 100-line MATLAB code, $\mathtt{TOPress}$, for topology optimization of structures subjected to fluidic pressure loads using the method of moving asymptotes. The code is intended for pedagogical purposes and aims to ease the beginners' and students' learning toward the topology optimization with design-dependent fluidic pressure loads. \texttt{TOPress} is developed per the approach first reported in~Kumar et al. (Struct Multidisc Optim 61(4):1637–1655, 2020). The Darcy law, in conjunction with the drainage term, is used to model the applied pressure load. The consistent nodal loads are determined from the obtained pressure field. The employed approach facilitates inexpensive computation of the load sensitivities using the adjoint-variable method. Compliance minimization subject to volume constraint optimization problems are solved. The success and efficacy of the code are demonstrated by solving  benchmark numerical examples involving pressure loads, wherein the importance of load sensitivities is also demonstrated. \texttt{TOPress} contains six main parts, is described in detail, and is extended to solve different problems. Steps to include a projection filter are  provided to achieve loadbearing designs close to~0-1. The code is provided in Appendix~\ref{Sec:TOPress} and can also be downloaded along with its extensions from \url{https://github.com/PrabhatIn/TOPress}. \\
	

	\vspace{-4mm}
	\rule{\linewidth}{.15mm}

\section{Introduction}
Nowadays, topology optimization (TO), a systematic design technique, is being used in almost all engineering fields~\citep{sigmund2013topology}. The technique provides an optimum material distribution for a given design problem involving certain single-/multi-physics concepts by extremizing the desired/formulated objective under a set of given physical/geometrical constraints. The design domain is described using finite elements (FEs) in a typical TO setting. Each element is assigned a design variable $\rho \in [0,\,1]$, which defines the material states of the element. $\rho = 0$ indicates void state of an element, whereas $\rho = 1$  represents its solid phase.

\begin{figure*}[h]
	\begin{subfigure}[t]{0.50\textwidth}
		\centering
		\includegraphics[scale=1]{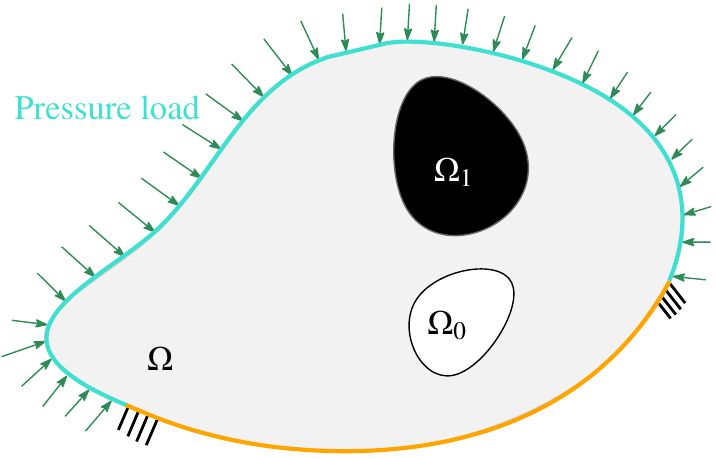}
		\caption{}
		\label{fig:Schematic1}
	\end{subfigure}
	\quad \quad
	\begin{subfigure}[t]{0.5\textwidth}
		\centering
		\includegraphics[scale=1]{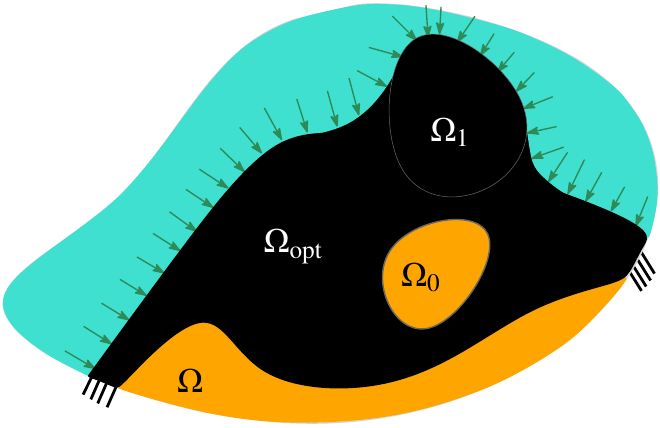}
		\caption{}
		\label{fig:Schematic2}
	\end{subfigure}
	\caption{A schematic diagram for a pressure-loaded problem. (\subref{fig:Schematic1}) Design domain. $\mathrm{\Omega}_1$ and $\mathrm{\Omega}_0$ denote the non-design solid and void regions, respectively. Fluidic pressure load is indicated via a set of arrows. Fixed boundary conditions are also depicted.  (\subref{fig:Schematic2}) A representative solution, $\mathrm{\Omega}_\text{opt}$} \label{fig:Schematic}
\end{figure*}
The applied input (external) load on structures can vary (design-dependent) or remain constant with optimization evolution. Though one encounters the former loads in various applications \citep{kumar2020topology}, handling them in a TO framework is a heavy and involved task as their location, magnitude, and/or direction alter with optimization progress. Thus, developing new approaches or using the existing ones may be challenging, resulting in stiff learning curves for beginners and students. In addition, no open education codes for TO with (fluidic) design-dependent pressure loads yet exist in the current state-of-the-art~\citep{wang2021comprehensive}. Therefore, the paper's primary aim is to offer a  pedagogical MATLAB code and its extensions for topology optimization of strucutres involving design-dependent fluidic pressure loads with all the required ingredients to reduce barriers to beginners' learning steps and fill the existing gap. And also, we except that the proposed code will provide a basic platform for developing and extending it for different applications involving pressure (pneumatic) loads, e.g., pneumatically activated soft robotics~\citep{xavier2022soft,kumar2023towards}, pressurized meta-material, etc. 

Providing educational codes on TO for pedagogical purposes is a welcomed trend that commenced with the celebrated 99-line MATLAB code written by \citet{sigmund200199}. An efficient version of 99-line code, \texttt{top88}, is presented in~\citet{andreassen2011efficient}. Codes use quadrilateral elements to parameterize the design domains. Such elements provide a node/point connection between two diagonally juxtaposed elements whereas polygonal/hexagonal elements offer non-singular connectivity. Therefore, by and large, geometrical anomalies (checkerboard patterns and point connections)  get subdued inherently with such FEs~\citep {saxena2011topology,talischi2012polytop,kumar2022honeytop90}. MATLAB codes with polygonal/hexagonal elements are also presented in~\citet{saxena2011topology,talischi2012polytop,kumar2022honeytop90}. Several MATLAB codes involving different physics also exist.  \citet{andreassen2014determine,xia2015design} present codes for material modeling using the homogenization method. \cite{gao2019concurrent} provide \texttt{ConTop2D} and \texttt{ConTop3D} MATLAB codes for TO of multiscale 2D and 3D composite structures, respectively.  An efficient  137-line TO MATLAB code is provided by \cite{han2021efficient} for the structures experiencing finite deformation using a bi-directional structural optimization method.  \citet{christiansen2021compact} present a MATLAB code for inverse design in phonotics. \citet{ferrari2021topology} provide a code in MATLAB for linearized buckling in TO. \cite{ali2022toward} present a MATLAB code for concurrent topology optimization for lightweight structures with high heat conductivity and stiffness. Readers can find a comprehensive list of educational papers in~\cite{wang2021comprehensive}.

A schematic diagram for the pressure load problem is depicted in Fig.~\ref{fig:Schematic1}. A set of arrows indicates the applied pressure load. In a representative optimized design solution (Fig.~\ref{fig:Schematic2}), one can see that the pressure load boundary shifts and settles within the design domain; thus, direction, and loading location alter. Therefore, such loads require a unique modeling technique in a TO framework.  \citet{hammer2000topology} were the first to model pressure load in a TO framework for designing loadbearing structures. Many approaches have been proposed after that; see \cite{picelli2019topology,kumar2020topology,kumar2021topology} for the comprehensive lists. In a typical TO environment with a design-dependent pressure load, an ideal approach is expected to provide suitable solutions to the following challenges (C$_i$): 
\begin{itemize}
	\item C$_1$: \textsl{How to relate the pressure load to the design variables.}
	\item C$_2$: \textsl{How to implicitly or explicitly locate bounding surface for applying the pressure load as TO advances}
	\item C$_3$: \textsl{How to convert the boundary pressure load (pressure field) to the consistent nodal forces.}
	\item C$_4$: \textsl{How to evaluate the load sensitivities efficiently that appear for the design-dependent forces in the sensitivity analysis.}
\end{itemize}
The method  in~\citet{kumar2020topology} employs the Darcy law with a drainage term to relate the pressure load and the design variables (C$_1$). That, in turn, also helps implicitly identify the pressure-bounding surface, i.e., a solution to C$_2$. The pressure load is consistently converted to nodal loads using a transformation matrix arising due to the equivalent body force, i.e., the approach takes care of C$_3$. Further, the method facilitates computationally cheap evaluation of the load sensitivities directly using the adjoint-variable method, i.e., it provides a viable and efficient solution to C$_4$. Note many previous TO approaches involving pressure loads neglect the load sensitivities that per~\cite{kumar2020topology} influence the optimized structures' shape and topology. The approach presented in~\citet{kumar2020topology} is also suitable for designing pressure-driven compliant mechanisms~\citep{kumar2020topology}, which is expected to open a potential direction towards developing soft robots using TO~\citep{kumar2023towards}. Further, the method is readily extended for optimizing 3D loadbearing designs and pressure-driven compliant mechanisms in~\cite{kumar2021topology}. Therefore, this paper adopts the approach herein for developing the proposed code.

This paper provides a compact 100-line MATLAB code, \texttt{TOPress}, for topology optimization of structures involving fluidic pressure loads using the method of moving asymptotes (MMA, cf.~\cite{svanberg1987}) for educational purposes. Though the MMA MATLAB code contains many lines~\citep{svanberg1987}, for brevity and compactness, the \texttt{mmasub} function call is counted as two lines in the provided 100-line code (Appendix~\ref{Sec:TOPress}). \texttt{TOPress} is developed per the approach proposed by~\cite{kumar2020topology}, and its various extensions are provided to solve benchmark problems. Steps to include a projection filter are also given to achieve close to black-and-white loadbearing structures. As mentioned above, the code uses the MMA (written in 1999 and updated in 2002 version)~\citep{svanberg1987} as the optimizer, which can permit users to readily extend the code for various applications involving fluidic pressure loads with multiple constraints and/or different additional physics. In addition, with a design-dependent load,  compliance objective may potentially lose its monotonous behavior with respect to the design variables due to load sensitivity terms, and, in such cases, the MMA optimizer is typically preferred~\citep{bruyneel2005note,kumar2022topology}.

The remainder of the paper is structured as follows. Sec.~\ref{Sec:Sec2} presents a topology optimization framework---pressure load modeling and nodal load calculation, demonstration of the Darcy law, and optimization problem formulation with sensitivity analysis. Sec.~\ref{Sec:Sec3} describes MATLAB implementation in detail and provides various extension for benchmark numerical examples. An extension of the 100-line code with the Heaviside projection filter is also provided. Lastly, concluding remarks are noted in Sec.~\ref{Sec:Sec4}. 

\section{Topology optimization framework}\label{Sec:Sec2}
The section describes the modeling of the pressure load, evaluation of the consistent nodal loads, objective formulation, and sensitivity analysis in short. One can refer to~\cite{kumar2020topology} for a detailed description.  
\subsection{Pressure load modeling and nodal loads evaluation}\label{Sec:pressMod}
As TO advances, the material states of FEs change. Thus, elements can be considered a porous medium at the beginning of TO. In addition, a pressure difference across the design domain is known already from the given pressure load boundary conditions. Therefore, the Darcy law determining the flux of a fluid flow due to pressure difference in a porous medium can be a natural and smart choice for modeling the fluidic pressure load \citep{kumar2020topology}. Mathematically, one writes the flux $\mathbf{q}$  using the available pressure gradient $\nabla p$, permeability $\kappa$  of the medium, and the fluid viscosity $\mu$ as

\begin{equation}\label{Eq:Darcyflux}
	\bm{q} = -\frac{\kappa}{\mu}\nabla p = -K(\tilde{\bm{\rho}}) \nabla p,
\end{equation}
where $\tilde{\bm{\rho}}$, the  physical design vector herein, is the filtered design vector \citep{bruns2001}; $K(\tilde{\bm{\rho}})$, the flow coefficient, is defined for element~$e$ using the flow contrast $\epsilon = \frac{K_s}{K_v}$ as
\begin{equation}\label{Eq:Flowcoefficient}
	K(\tilde{\rho_e}) = K_v\left(1-(1-\epsilon) \mathcal{H}(\tilde{{\rho_e}},\,\beta_\kappa,\,\eta_\kappa)\right),
\end{equation} 
where $K_v$ and $K_s$ are the flow coefficients of the void and solid phases of an element, and $\mathcal{H}(\bar{{\rho_e}},\,\beta_\kappa,\,\eta_\kappa) = \frac{\tanh{\left(\beta_\kappa\eta_\kappa\right)}+\tanh{\left(\beta_\kappa(\tilde{\rho}_e - \eta_\kappa)\right)}}{\tanh{\left(\beta_\kappa \eta_\kappa\right)}+\tanh{\left(\beta_\kappa(1 - \eta_\kappa)\right)}}$, is a smooth Heaviside function. $\{\eta_\kappa,\,\beta_\kappa\}$ are the flow parameters. $\eta_\kappa$ and $\beta_\kappa$ define the step position and slope of $K(\tilde{\rho_e})$, respectively. We set $K_v =1$, and $\epsilon = \SI{1e-7}{}$, i.e., $K_s = \epsilon$~\citep{kumar2021topology} in the provided code, \texttt{TOPress}. To this end, we write $K(\tilde{\rho_e}) = 1-(1-\epsilon) \mathcal{H}(\tilde{{\rho_e}},\,\beta_\kappa,\,\eta_\kappa)$. Using the fundamentals of the state equilibrium, the balanced equation corresponding to~\eqref{Eq:Darcyflux} is
\begin{equation}\label{Eq:DarcyFEM}
	\nabla \cdot \bm{q} = -\nabla \cdot \left(K(\tilde{\bm{\rho}}) \nabla p\right) = 0
\end{equation}

\begin{figure*}[h!]
	\begin{subfigure}[t]{0.3\textwidth}
		\centering
		\includegraphics[scale=0.85]{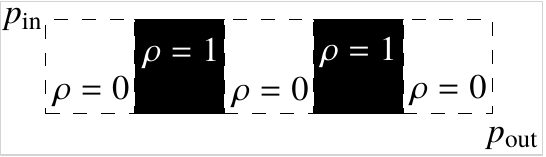}
		\caption{}
		\label{fig:DarcyDD}
	\end{subfigure}
	\quad \quad
	\begin{subfigure}[t]{0.3\textwidth}
		\centering
		\includegraphics[scale=0.85]{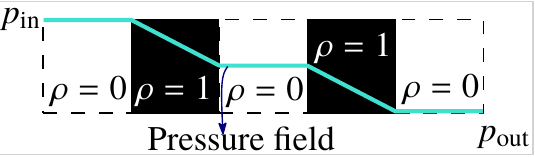}
		\caption{}
		\label{fig:DarcyPF}
	\end{subfigure}
	\quad \quad
	\begin{subfigure}[t]{0.3\textwidth}
		\centering
		\includegraphics[scale=0.85]{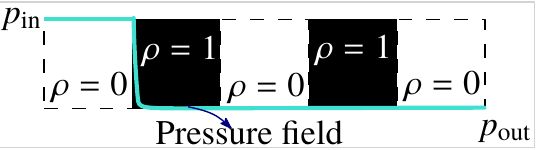}
		\caption{}
		\label{fig:DarcyAPF}
	\end{subfigure}
	\caption{(\subref{fig:DarcyDD}) A test design domain, $p_\text{in}$ and $p_\text{out}$ are the input and output fluidic pressure, respectively. (\subref{fig:DarcyPF}) The obtained pressure field variation from the solution of the Darcy law (Eq.~\ref{Eq:DarcyFEM}), and (\subref{fig:DarcyAPF}) The desired pressure field variation.} \label{fig:solution}
\end{figure*}

\begin{figure}[h] 
	\begin{subfigure}{0.3\textwidth}
		\centering
		\includegraphics[scale=1]{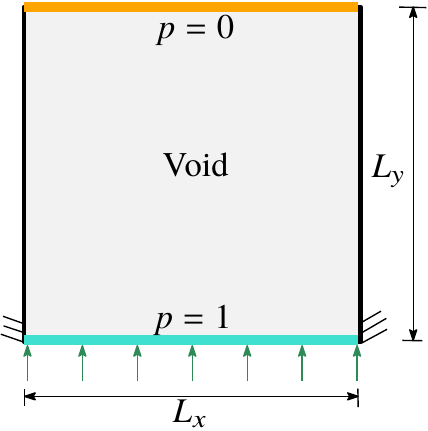}
		\caption{SP1}
		\label{fig:DD1}
	\end{subfigure}
	\quad 
	\begin{subfigure}{0.3\textwidth}
		\centering
		\includegraphics[scale=1]{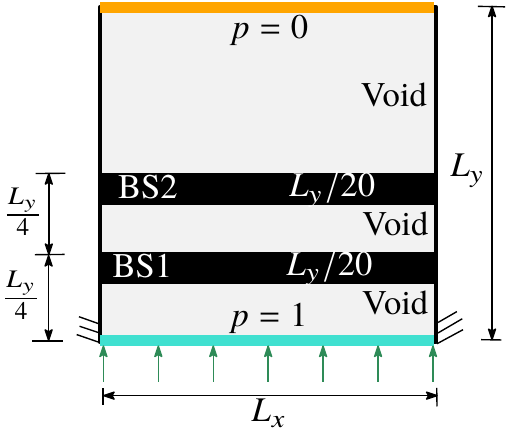}
		\caption{SP2}
		\label{fig:DD2}
	\end{subfigure}
	\quad
	\begin{subfigure}{0.3\textwidth}
		\centering
		\includegraphics[scale=1]{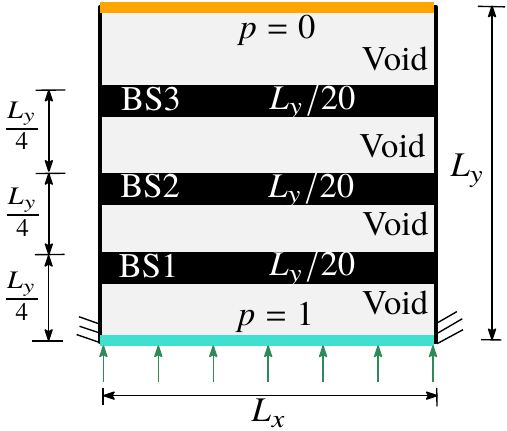}
		\caption{SP3}
		\label{fig:DD3}
	\end{subfigure}
	\caption{Three sample problems (SP1,\,SP2,\,SP3) demonstrate the working of the Darcy law. $L_x\times L_y = \SI{200}{} \times \SI{200}{}$. Pressure loads of $\SI{1}{}$ and $\SI{0}{}$ are applied on the bottom and top edges, respectively. Strips in black with width $\frac{L_y}{20}$ are with $\rho =1$.} \label{fig:ExampleDD}
\end{figure}
\begin{figure}[h]
	\centering
	\includegraphics[scale = 0.5]{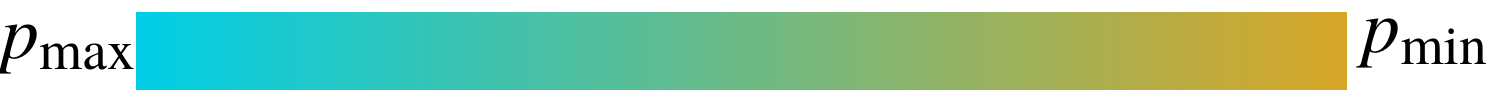}
	\caption{Pressure colorbar. ${p}_\text{max} = \SI{1}{}$ and $p_\text{min} = \SI{0}{}$.}
	\label{fig:pressurecolorbar}
\end{figure}

\begin{figure}[h] 
	\begin{subfigure}{0.22\textwidth}
		\centering
		\includegraphics[scale=0.30]{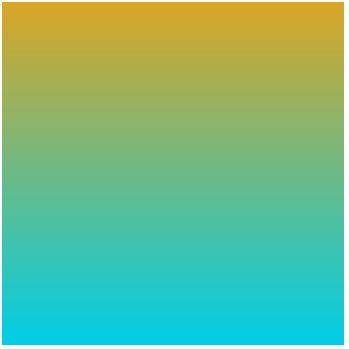}
		\caption{SP1 P-field}
		with \texttt{DrainT}=0 or 1
		\label{fig:SP1_Press}
	\end{subfigure}
	\begin{subfigure}{0.18\textwidth}
		\centering
		\includegraphics[scale=0.30]{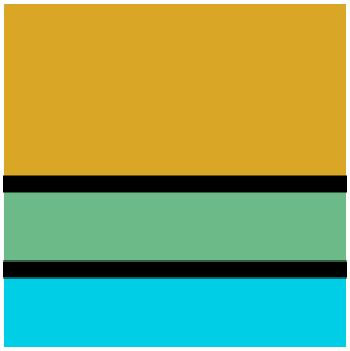}
		\caption{SP2 P-field }
		with \texttt{DrainT}=0
		\label{fig:SP2_Press_DraintT_0}
	\end{subfigure}
	\begin{subfigure}{0.18\textwidth}
		\centering
		\includegraphics[scale=0.30]{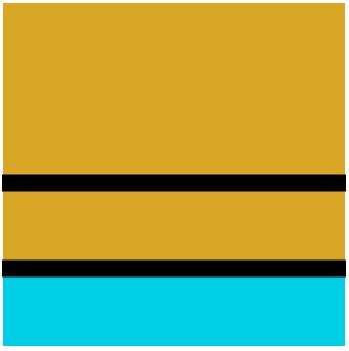}
		\caption{SP2 P-field }
		with \texttt{DrainT}=1
		\label{fig:SP2_Press_DraintT_1}
	\end{subfigure}
	\begin{subfigure}{0.18\textwidth}
		\centering
		\includegraphics[scale=0.30]{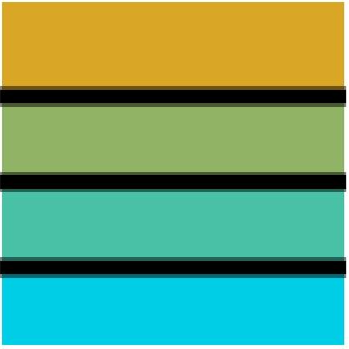}
		\caption{SP3 P-field }
		with \texttt{DrainT}=0
		\label{fig:SP3_Press_DraintT_0}
	\end{subfigure}
	\begin{subfigure}{0.18\textwidth}
		\centering
		\includegraphics[scale=0.30]{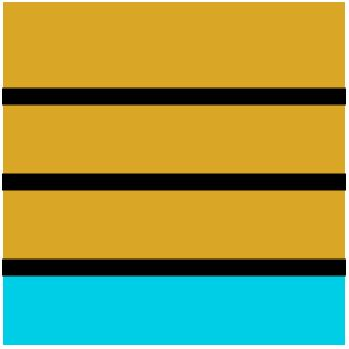}
		\caption{SP3 P-field }
		with \texttt{DrainT}=1
		\label{fig:SP3_Press_DraintT_1}
	\end{subfigure}
	\caption{Pressure field (P-field) plots.} \label{fig:ExamplePressure_plot}
\end{figure}

\begin{figure}[h] 
	\begin{subfigure}{0.48\textwidth}
		\centering
		\begin{tikzpicture} 	
			\pgfplotsset{compat =1.9}
			\begin{axis}[
				width = 1\textwidth,
				xlabel= $y-$coordinates,
				ylabel= Pressure,
				legend style={at={(0.85,0.80)},anchor=east}]
				\pgfplotstableread{SP2_Press_DrainT_0.txt}\mydata;
				\addplot[red, mark size=1pt,style={very thick}]
				table {\mydata};
				\addlegendentry{\texttt{DrainT}=0}
				\pgfplotstableread{SP2_Press_DrainT_1.txt}\mydata;
				\addplot[blue,mark = *,mark size=2pt,style= thick]
				table {\mydata};
				\addlegendentry{\texttt{DrainT}=1}
			\end{axis}
		\end{tikzpicture}
		\caption{P-field}
		\label{fig:SP2_Press_comp}
	\end{subfigure}
	\begin{subfigure}{0.48\textwidth}
		\centering
		\begin{tikzpicture} 	
			\pgfplotsset{compat =1.9}
			\begin{axis}[
				width = 1\textwidth,
				xlabel= $y-$coordinates,
				ylabel= Nodal force,
				legend style={at={(0.85,0.80)},anchor=east}]
				\pgfplotstableread{SP2_force_DrainT_0.txt}\mydata;
				\addplot[red, mark size=2pt,style={very thick}]
				table {\mydata};
				\addlegendentry{\texttt{DrainT}=0}
				\pgfplotstableread{SP2_force_DrainT_1.txt}\mydata;
				\addplot[blue,mark = *,mark size=2pt,style= thick]
				table {\mydata};
				\addlegendentry{\texttt{DrainT}=1}
			\end{axis}
		\end{tikzpicture}
		\caption{Nodal forces}
		\label{fig:SP2_force_comp}
	\end{subfigure}
	\caption{P-field and nodal forces along the vertical central line for SP2.} \label{fig:SP2_Press_force_comp}
\end{figure}
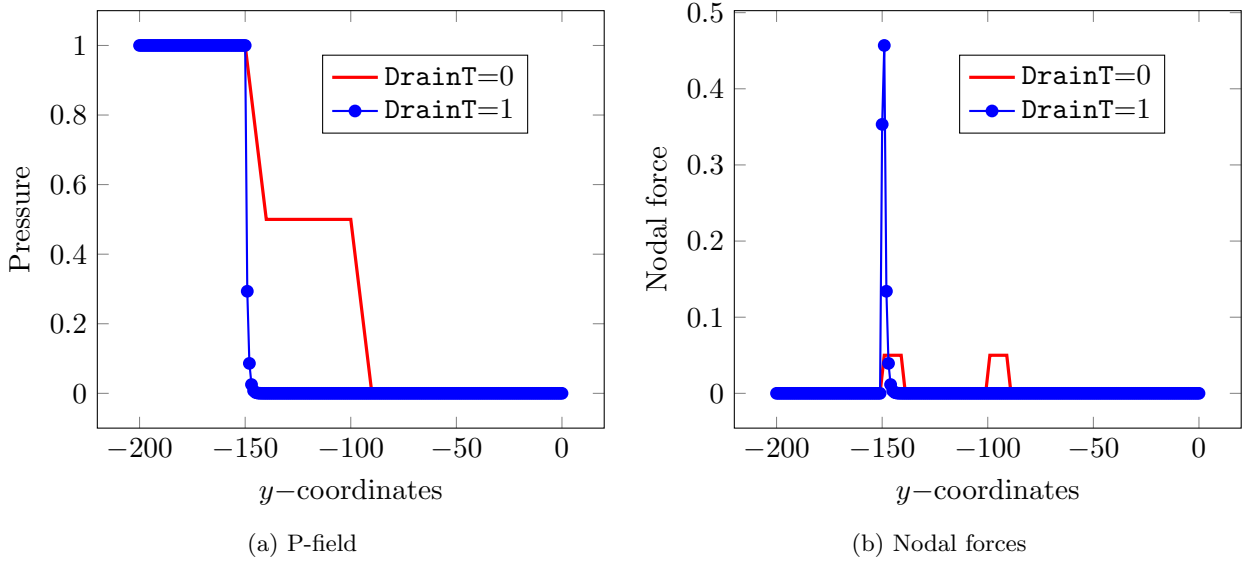

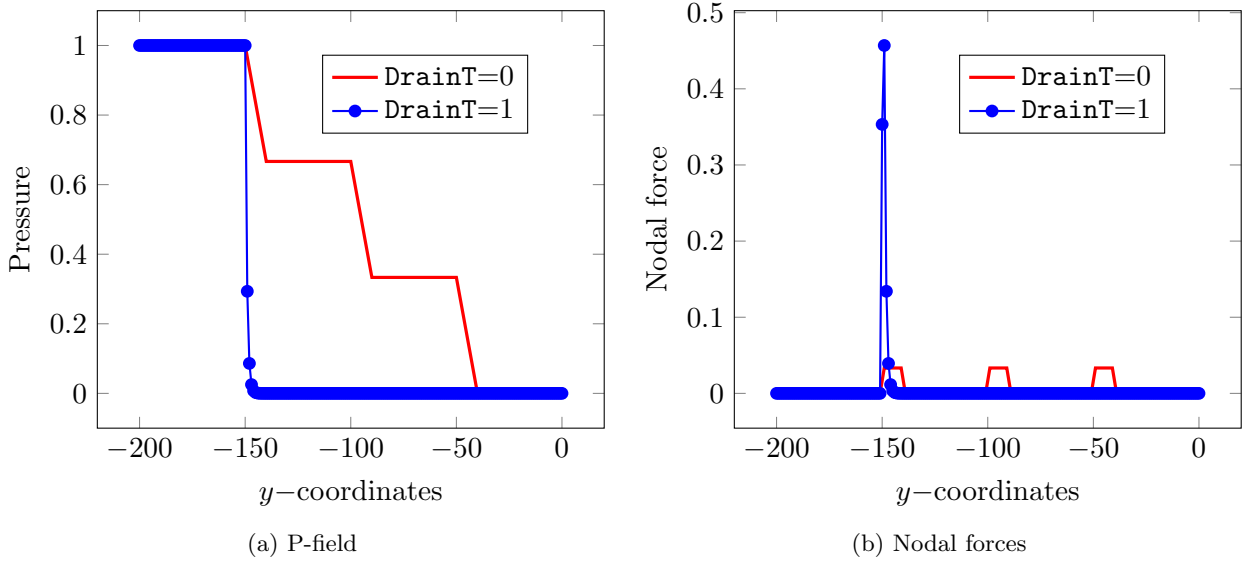
\begin{figure}[h] 
	\begin{subfigure}{0.48\textwidth}
		\centering
		\begin{tikzpicture} 	
			\pgfplotsset{compat =1.9}
			\begin{axis}[
				width = 1\textwidth,
				xlabel= $y-$coordinates,
				ylabel= Pressure,
				legend style={at={(0.85,0.80)},anchor=east}]
				\pgfplotstableread{SP3_Press_DrainT_0.txt}\mydata;
				\addplot[red, mark size=1pt,style={very thick}]
				table {\mydata};
				\addlegendentry{\texttt{DrainT}=0}
				\pgfplotstableread{SP3_Press_DrainT_1.txt}\mydata;
				\addplot[blue,mark = *,mark size=2pt,style= thick]
				table {\mydata};
				\addlegendentry{\texttt{DrainT}=1}
			\end{axis}
		\end{tikzpicture}
		\caption{P-field}
		\label{fig:SP3_Press_comp}
	\end{subfigure}
	\begin{subfigure}{0.48\textwidth}
		\centering
		\begin{tikzpicture} 	
			\pgfplotsset{compat =1.9}
			\begin{axis}[
				width = 1\textwidth,
				xlabel= $y-$coordinates,
				ylabel= Nodal force,
				legend style={at={(0.85,0.80)},anchor=east}]
				\pgfplotstableread{SP3_force_DrainT_0.txt}\mydata;
				\addplot[red, mark size=2pt,style={very thick}]
				table {\mydata};
				\addlegendentry{\texttt{DrainT}=0}
				\pgfplotstableread{SP3_force_DrainT_1.txt}\mydata;
				\addplot[blue,mark = *,mark size=2pt,style= thick]
				table {\mydata};
				\addlegendentry{\texttt{DrainT}=1}
			\end{axis}
		\end{tikzpicture}
		\caption{Nodal forces}
		\label{fig:SP3_force_comp}
	\end{subfigure}
	\caption{P-field and nodal forces along the vertical central line for SP3.} \label{fig:SP3_Press_force_comp}
\end{figure}
\noindent  One solves \eqref{Eq:DarcyFEM} using the standard finite element method to determine the pressure field across the given design domain. The given design domain is depicted in Fig.~\ref{fig:DarcyDD}. $p_\text{in}$ and $p_\text{out}=0$ are the input and output fluidic pressure, respectively. With the Darcy law solution~\eqref{Eq:DarcyFEM} and the given pressure load boundary conditions,  the obtained pressure field for this problem (Fig.~\ref{fig:DarcyDD}) is as displayed in Fig.~\ref{fig:DarcyPF}, which is physically unrealistic as the pressure gradient exists across the solid regions ($\rho=1$). Thus, to achieve the natural and desired pressure field/distribution,  a conceptualized volumetric drainage per second in a unit volume term, $Q_\text{drain}$, with the Darcy law is used~\citep{kumar2020topology}. $Q_\text{drain}$ acts as a pressure drainer from the solid elements. The final balance equation in conjunction with $Q_\text{drain}$ can be written as \citep{kumar2020topology}
\begin{equation}\label{Eq:stateequation}
	\nabla\cdot\bm{q} - Q_\text{drain} = \nabla \cdot \left(K(\tilde{\bm{\rho}}) \nabla p\right)+  Q_\text{drain}=0.
\end{equation}
${Q}_\text{drain} = -D(\bar{\rho_e}) (p - p_{\text{ext}})$ with $D(\bar{\rho_e}) =  D_{\text{s}}\mathcal{H}(\bar{{\rho_e}},\,\beta_d,\,\eta_d)$. $\left\{\eta_\text{d},\,\beta_\text{d}\right\}$ are the drainage parameters. Note, $D_{\text{s}} = \left(\frac{\ln{r}}{\Delta s}\right)^2 K_\text{s}$, where $r = \frac{p|_{\Delta s}}{p_\text{in}}$; $\Delta s$, a penetration parameter, is set to width/height of a few FEs, and $p|_{\Delta s}$ is the pressure at $\Delta s$. Based on our experience,  $r\in[0.001 \,\,  0.1]$. For all practical purposes and to reduce the number of user-defined parameters, we consider $\eta_h = \eta_k = \eta_f$ and $\beta_h = \beta_k =\beta_f$ in the provided code \texttt{TOPress}. However, one can choose them differently per the recommendation outlined in~\cite{kumar2020topology}. In a discrete setting, using the standard finite element steps, \eqref{Eq:stateequation}, for element~$i$, yields  considering $p_\text{out} = 0$ and $\mathbf{q}_\mathrm{\Gamma} =0$ to~\citep{kumar2020topology}

\begin{equation} \label{Eq:FEAformulation}
	\begin{aligned}
		&\left[\int_{\Omega_e}\left( K~ \mathbf{B}^\top_\text{p} \mathbf{B}_\text{p}\right) \text{d} V  + \int_{\Omega_e}\left( D ~\mathbf{N}^\top_\text{p} \mathbf{N}_\text{p}\right) \text{d} V~\right]\mathbf{p}_e = \mathbf{0},
	\end{aligned}
\end{equation}
where $\mathbf{N}_\text{p} = [N_1,\,N_2,\,N_3,\,N_4]^\top$ are the bilinear shape functions for the quadrilateral elements employed to parameterized the design domains, $\mathbf{B}^\top_\text{p} = \nabla\mathbf{N}_\text{p}$, and $\mathbf{p}_e = [p_1,\,p_2,\,p_3,\,p_4]^\top$. Note $N_i$ and $p_i$ ($i = 1,\,\cdots,\,4$) represent the shape function and pressure degree of freedom for the $i^\text{th}$ node of element~$e$, respectively (see Fig.~\ref{fig:nomenclatureSigle}).
In light of numerical integration, \eqref{Eq:FEAformulation} yields to
\begin{equation} \label{Eq:FEAnumint}
	\begin{aligned}
		\mathbf{K}_\text{p}^e \mathbf{p}_e+ \mathbf{K}^e_\text{Dp} \mathbf{p}_e = \mathbf{A}_e \mathbf{p}_e = \mathbf{0},
	\end{aligned}
\end{equation}
where $\mathbf{\mathbf{K}}_\text{p}^e$ and $\mathbf{K}^e_\text{Dp}$ are the element flow matrices due to the Darcy law and the drainage term, respectively (See Appendix~\ref{Sec:APPExpression}). $\mathbf{A}_e$ indicates the overall element flow matrix. At the global level, \eqref{Eq:FEAnumint} transpires to
\begin{equation}\label{Eq:PDEsolutionpressure}
	\mathbf{Ap} = \mathbf{0},
\end{equation}
where $\mathbf{p}$ and $\mathbf{A}$ are the global pressure vector and flow matrix, respectively. Note, $\mathbf{A}$ is obtained by assembling $\mathbf{K}_\text{p}^i$ and $\mathbf{K}^i_\text{Dp}$. The solution to \eqref{Eq:PDEsolutionpressure}  gives the pressure field $\mathbf{p}$ as a function of the design variables. To this end, when the problem shown in Fig.~\ref{fig:DarcyDD} is solved using \eqref{Eq:PDEsolutionpressure}, one obtains the pressure field as indicated in Fig.~\ref{fig:DarcyAPF}, which is indeed a realistic pressure distribution.

Further, the obtained $\mathbf{p}$ is expressed as an equivalent body force  using the balance equation, which gives \citep{kumar2020topology}
\begin{equation}\label{Eq:pressuretoload}
	\bm{b} \text{d}V = -\nabla p \text{d}V,
\end{equation}
where $\bm{b}$ is the body force per unit volume. In view of the standard finite element methods, \eqref{Eq:pressuretoload} provides
\begin{equation}\label{Eq:Forcepressureconversion}
	\mathbf{F}_e = -\left[\int_{\mathrm{\Omega}_e} \trr{\mathbf{N}}_\mathbf{u} \mathbf{B}_p  d {V}\right]\, \mathbf{p}_e = \mathbf{T}_e\,\mathbf{p}_e,
\end{equation}
where $\mathbf{N}_\mathbf{u} = [N_1\mathbf{I},\, N_2\mathbf{I},\,N_3\mathbf{I},\,N_4\mathbf{I}]$, with $\mathbf{I}$  the identity matrix in $\mathcal{R}^2$. In global sense, \eqref{Eq:pressuretoload} yields to
\begin{equation}\label{Eq:nodalforce}
	\mathbf{F} = -\mathbf{T}\mathbf{p},
\end{equation}
where $\mathbf{T}$ is the global transformation matrix, and $\mathbf{F}$ is the global force vector.

To summarize, \eqref{Eq:PDEsolutionpressure} is used for determining the pressure field, whereas the corresponding consistent nodal force vector  for an element is calculated using \eqref{Eq:Forcepressureconversion}.

\subsection{Demonstration of the Darcy law}
To demonstrate  success of the Darcy law with drainage term, we consider a set of three problems, SP1, SP2, and SP3, shown in Fig.~\ref{fig:ExampleDD}. SP1 is considered fully void, whereas SP2 and SP3 contain two and three solid bars of width $\frac{L_y}{20}$ each. We use $L_x$ and $L_y$ to represent the dimensions of the designs in $x-$ and $y-$directions, respectively. $L_x = \SI{200}{}$ and $L_y = \SI{200}{}$ are set. The out of plane thickness, $\texttt{t}$, is set to \SI{1}{}. The bottom edges of the domains experience $\texttt{Pin}=\SI{1}{}$ pressure load, whereas their top edges are placed on zero pressure load. The bottom parts of the left and right edges are fixed. The domains are parameterized by $\texttt{nelx}\times \texttt{nely} = \texttt{nel}=200 \times 200$ square bi-linear finite elements, where $\texttt{nelx}$ and $\texttt{nely}$ indicate number of the squared finite elements (FEs) in $x-$ and $y-$directions, respectively. \texttt{nel} is the total number of finite elements used to parameterize the design domains. Fig.~\ref{fig:pressurecolorbar} provides the color scheme used to display the pressure variation for the problems. 

Figure~\ref{fig:ExamplePressure_plot} shows the pressure field (P-field) variation for SP1, SP2, and SP3. For SP2 and SP3, two cases are considered--with drainage term (\texttt{DrainT=1}) and without drainage term (\texttt{DrainT=0}). Fig.~\ref{fig:SP2_Press_DraintT_0}  and Fig.~\ref{fig:SP3_Press_DraintT_0} display pressure variation for SP2 and SP3 with \texttt{DrainT=0}, respectively . They are unrealistic pressure variations, as one can observe a pressure gradient across the solid strips. A realistic pressure distribution (no pressure gradient across the solid bars) can be noted with \texttt{DrainT}=1 (Fig.~\ref{fig:SP2_Press_DraintT_1} and Fig.~\ref{fig:SP3_Press_DraintT_1}). The pressure field (P-field) and nodal forces along the vertical central line are plotted in Fig.~\ref{fig:SP2_Press_force_comp} for SP2. One again notes an unrealistic pressure field with \texttt{DrainT}=0 (Fig.~\ref{fig:SP3_Press_DraintT_0}). The variation of pressure and nodal forces along the central line of SP3 are displayed in Fig.~\ref{fig:SP3_Press_comp} and Fig.~\ref{fig:SP3_force_comp}, respectively.  For all the sample problems, we find  $\texttt{MFx} = \SI{0.0}{}$ and $\texttt{MFy} =\SI{200.0}{}$. $x-$ and $y-$ magnitudes of the forces are indicated by \texttt{MFx} and \texttt{MFy}, respectively. The force applied by the given pressure load is $\texttt{Pin}\times \texttt{t} \times {L_x} = 1\times 1 \times 200 = \SI{200}{}$ in the vertical direction, which equals to \texttt{MFy}, and that indeed proves the correctness of the pressure load modeling. Therefore, herein it gets established that the Darcy formulation with the drainage term ($\texttt{DrainT} = 1$) gives the realistic pressure field while maintaining the force magnitude. With regard to this study, \texttt{TOPress} uses \texttt{DraintT=1}, as default. Next, we present the topology optimization formulation for the compliance minimization problem with the above-described Darcy law.

\subsection{Topology optimization formulation}\label{Sec:TopOptFor}
We present \texttt{TOPress} code for the compliance minimization with the given volume fraction for the loadbearing structures. The optimization problem can be written as
\begin{equation} \label{Eq:OPTI} 
	\begin{rcases}
		\begin{split}
			&{\min_{\tilde{\bm{\rho}}}} \quad C({\tilde{\bm{\rho}}}) = \mathbf{u}^\top \mathbf{K}(\tilde{\bm{\rho}})\mathbf{u} = \sum_{j=1}^{\mathtt{nel}}\mathbf{u}_j^\top\mathbf{k}_j(\rho_j)\mathbf{u}_j\\
			&\text{subjected to:}\\
			&\bm{\lambda}_1:\,\,\mathbf{A} \mathbf{p} = \mathbf{0}\\
			&\bm{\lambda}_2:\,\,\mathbf{K} \mathbf{u} = \mathbf{F} = -\mathbf{T}\mathbf{p}\\
			&\Lambda:V(\tilde{\bm{\rho}})-V^* \le 0\\
			&\quad\,\,\,\, \bm{0} \leq \bm{\rho} \leq \bm{1} \\
			&\text{Data:} \quad V^*,\,E_0,\,E_\text{min},\,p,\,K_v,\,\epsilon
		\end{split}
	\end{rcases},
\end{equation} 
where $C$ indicates the structure's compliance and \texttt{nel} is the total number of elements used to describe the design domain. $\mathbf{u}$ and $\mathbf{K}$ represent the global displacement vector and stiffness matrix, respectively. $V^*$ indicates the permitted resource, and $V$ represents the current volume of the evolving domain. $\mathbf{F}$, the global force vector, appears due to the pressure loads. $\bm{\lambda}_1$ (vector), $\bm{\lambda}_2$ (vector) and $\Lambda$ (scalar) are the Lagrange multipliers. $\tilde{\bm{\rho}}$ is the filtered design vector corresponding to the actual design variable vector $\bm{\rho}$. We use the classical density filter~\citep{bruns2001} in the provided codes. The filtered design variable for element $i$ is determined as:
\begin{equation} \label{EQ:density_filter}
	\tilde{\rho}_i = \frac{\displaystyle\sum_{j=1}^{\texttt{nel}}\rho_j \mathrm{v}_j \mathrm{w}(\mathbf{x}_i,\mathbf{x}_j)}{\displaystyle\sum_{j=1}^{\texttt{nel}} \mathrm{v}_j \mathrm{w}(\mathbf{x}_i,\mathbf{x}_j) } \; ,
\end{equation}
where  weight $\mathrm{w}(\mathbf{x}_i,\mathbf{x}_j)$ = $\mathrm{max}  \left(0 \; , \; 1-\frac{\| \mathrm{\mathbf{x}}_i - \mathrm{\mathbf{x}}_j \|}{\mathrm{r}_\mathrm{fill}} \right)$ \citep{bruns2001}. $r_\text{fill}$ is the filter radius. $\mathrm{v}_j$ is the volume of element $j$. Note that $\mathrm{v}_j$ and $\mathrm{w}(\mathrm{x}_i,\mathrm{x}_j)$ do not alter with TO iterations; thus, they are determined once at the beginning of the optimization and are stored in a matrix $\mathbf{H}$ as:
\begin{equation}
	\mathrm{H}_{i,j} = \frac{\mathrm{v}_j \: \mathrm{w}(\mathrm{\mathbf{x}}_i,\mathrm{\mathbf{x}}_j)}{{\displaystyle\sum_{j=1}^{\texttt{nel}} \mathrm{v}_j \mathrm{w}(\mathrm{\mathbf{x}}_k,\mathrm{\mathbf{x}}_j) }} \: .
\end{equation} 
To this end, one writes the filtered design vector as $\bm{\tilde{\rho}}=\mathbf{H}\bm{\rho}$ and its derivative as $\frac{\partial \bm{\tilde{\rho}}}{\partial \bm{\rho}}=\mathbf{H}^\top$. We perform the filtering using \texttt{imfilter} MATLAB function in the 100-line code (see Appendix B). Note that one may also use sensitivity filtering as per~\cite{sigmund200199}. 

At each TO iteration, $\mathbf{u}$ is determined by solving the structure state equation,  $\mathbf{K} \mathbf{u} = \mathbf{F} = -\mathbf{T}\mathbf{p}$, wherein $\mathbf{p}$ is the solve of fluid flow equilibrium $\mathbf{Ap} = \mathbf{0}$, and $\mathbf{T}$ is the transformation matrix as described above. Stiffness matrix $\mathbf{K}$ depends upon the physical design variables wherein the modified solid Isotropic Material with Penalization (SIMP) scheme \citep{sigmund2007} is employed to interpolate  Young's modulus of element $i$. Mathematically,
\begin{equation}
	\mathrm{E}_i = \mathrm{E}_\mathrm{min} + \tilde{\rho}^p_i(\mathrm{E}_0-\mathrm{E}_\mathrm{min}),
\end{equation}
where $p$, the SIMP parameter encourages convergence of TO towards 0-1 solutions. $E_0$ and $E_\text{min}$ are Young's moduli of the solid and void states of an element, respectively, and $E_i$ is Young's modulus of element~$i$.

\subsubsection{Sensitivity analysis}
The presented 100-line code and its extensions employ a gradient-based optimizer, the method of moving asymptotes (MMA, cf. \cite{svanberg1987}), to solve the optimization problem \eqref{Eq:OPTI}. Therefore, derivatives of the objective and constraints with respect to design variables are required. We use the adjoint-variable method  to determine the sensitivities by evaluating the Lagrange multipliers corresponding to the balance equations. Typically, one uses the adjoint-variable method wherein the number of design variables exceeds the number of constraints/state-dependent responses. The augmented performance function $\mathcal{L}$ is defined using the objective function and equilibrium equations \eqref{Eq:OPTI} as

\begin{equation}\label{Eq:Lagrange}
	\mathcal{L} = \mathbf{u}^\top \mathbf{K}\mathbf{u} + \bm{\lambda}_1^\top \mathbf{AP} + \bm{\lambda}_2^\top \left(\mathbf{KU + TP}\right).
\end{equation} 
Differentiating \eqref{Eq:Lagrange} with respect to physical variables, that yields
\begin{equation}\label{Eq:Lagrangeder}
	\begin{split}
		\frac{d \mathcal{L}}{d \tilde{\bm{\rho}}} = &2\mathbf{u}^\top\mathbf{K}\frac{\partial\mathbf{u}}{\partial \tilde{\bm{\rho}}} + \mathbf{u}^\top \frac{\partial\mathbf{K}}{\partial \tilde{\bm{\rho}}}\mathbf{u} + \bm{\lambda}_1^\top \left(\frac{\partial\mathbf{A}}{\partial\tilde{\bm{\rho}}}\mathbf{p}\right) + \bm{\lambda}_1^\top\left(\mathbf{A}\frac{\partial\mathbf{p}}{\partial\tilde{\bm{\rho}}}\right)\\ &+ \bm{\lambda}_2^\top \left(\frac{\partial\mathbf{K}}{\partial \tilde{\bm{\rho}}}\mathbf{u} + \mathbf{K}\frac{\partial\mathbf{u}}{\partial \tilde{\bm{\rho}}}\right) + \bm{\lambda}_2^\top \left(\frac{\partial\mathbf{T}}{\partial \tilde{\bm{\rho}}}\mathbf{p} + \mathbf{T}\frac{\partial\mathbf{p}}{\partial \tilde{\bm{\rho}}}\right)\\
		=& \mathbf{u}^\top \frac{\partial\mathbf{K}}{\partial \tilde{\bm{\rho}}}\mathbf{u} + \bm{\lambda}_2^\top \left(\frac{\partial\mathbf{K}}{\partial \tilde{\bm{\rho}}}\mathbf{u}\right) + \bm{\lambda}_1^\top \left(\frac{\partial\mathbf{A}}{\partial\tilde{\bm{\rho}}}\mathbf{p}\right) \\ &+ \underbrace{\left(2\mathbf{u}^\top\mathbf{K} + \bm{\lambda}_2^\top \mathbf{K}\right)}_{\Theta_1}\frac{\partial\mathbf{u}}{\partial \tilde{\bm{\rho}}} + \underbrace{\left(\bm{\lambda}_1^\top\mathbf{A} + \bm{\lambda}_2^\top \mathbf{T} \right)}_{\Theta_2}\frac{\partial\mathbf{p}}{\partial\tilde{\bm{\rho}}}
	\end{split}
\end{equation} 
In light of the fundamentals of the adjoint-variable method, we select $\bm{\lambda}_1$ and  $\bm{\lambda}_2$ such that $\Theta_1 = 0$ and $\Theta_2 = 0$, which give
\begin{equation} \label{Eq:LagrangeMultiplier}
	\begin{aligned}
		\bm{\lambda}_2 &= -2\mathbf{u}, \\
		\bm{\lambda}_1^\top &= - 	\bm{\lambda}_2^\top \mathbf{T} \mathbf{A}^{-1} = 2\mathbf{u}^\top \mathbf{T} \mathbf{A}^{-1},
	\end{aligned}
\end{equation}
and
\begin{equation}\label{Eq:senstivities_Obj}
	\frac{d {C}}{d \tilde{\bm{\rho}}} = -\mathbf{u}^\top \frac{\partial\mathbf{K}}{\partial \tilde{\bm{\rho}}}\mathbf{u} + \underbrace{2\mathbf{u}^\top \mathbf{T} \mathbf{A}^{-1}\frac{\partial\mathbf{A}}{\partial\tilde{\bm{\rho}}}\mathbf{p}}_{\text{Load sensitivities}}
\end{equation}
One notes, the load sensitivities modify the total sensitivities of compliance. With a constant load case, the compliance sensitivities contain only the first term of \eqref{Eq:senstivities_Obj}. The load sensitivity terms affect the monotonous behavior of the compliance objective of the design problem with the design-dependent loads~\citep{kumar2022topology}. Finally, 
one can use the chain rule to determine the derivatives of the objective with respect to the design variables as
\begin{equation}\label{Eq:objderivative}
	\frac{d {C}}{d \bm{\rho}} = \frac{d {C}}{d \tilde{\bm{\rho}}} \frac{d \tilde{\bm{\rho}}}{d \bm{\rho}}
\end{equation}
The derivative of the volume constraint is calculated per \citet{sigmund200199}, which is straightforward.   Having discussed all the ingredients needed for the pressure load formulation in a density-based TO formulation, next, we present the MATLAB implementation for the 100-line code and its various extensions for different pressure-loaded problems.   

\section{MATLAB Implementation}\label{Sec:Sec3}
\begin{figure}[h]
	\centering
	\includegraphics[scale = 1]{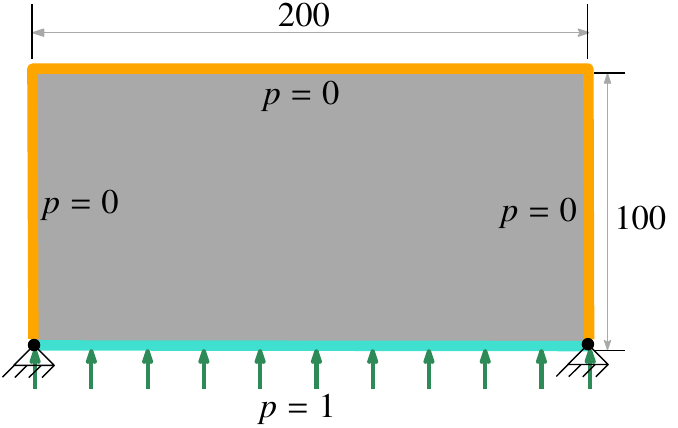}
	\caption{Internally pressurized arch design. Pressure load is applied on the bottom edges, whereas remaining edges are fixed at 0 pressure.}
	\label{fig:2Darcproblem}
\end{figure}

The internally pressurized arch design is the classical problem in TO with design-dependent pressure load. The problem was first presented and solved by~\cite{hammer2000topology}. Many researchers also consider the problem to test their method after that, e.g., \cite{chen2001topology,du2004topological,sigmund2007topology,zhang2008new,xia2015topology,picelli2015bi,emmendoerfer2018level,picelli2019topology,neofytou2020level,kumar2020topology,ibhadode2020topology,huang2022thermal}, to name a few. Thus, we consider the arc problem to demonstrate the presented 100-line MATLAB code, \texttt{TOPress}. The design domain, pressure, and structure boundary conditions for the internally pressurized arch are displayed in Fig.~\ref{fig:2Darcproblem}.

One calls \texttt{TOPress} from the MATLAB prompt as
\begin{lstlisting}[basicstyle=\scriptsize\ttfamily,breaklines=true,numbers=none,frame=tb]
	TOPress(nelx,nely,volf,penal,rmin,etaf,betaf,lst,maxit)
\end{lstlisting}
where \texttt{volf} is the permitted volume fraction, \texttt{penal} is the SIMP penalization parameter $p$, \texttt{rmin} is the filter radius $r_\text{fill}$, \texttt{etaf} denotes both (the step position) $\eta_\kappa$ and $\eta_d$ (Sec.~\ref{Sec:pressMod}), \texttt{betaf} indicates $\beta_\kappa$ and $\beta_d$ (Sec.~\ref{Sec:pressMod}) and \texttt{maxit} is the maximum number for the MMA optimizer. \texttt{lst} determines participation of load sensitivities in (\ref{Eq:senstivities_Obj}), i.e., $\texttt{lst}=1$ indicates load sensitivities are included, whereas $\texttt{lst}=0$ implies otherwise. 

\texttt{TOPress} primarily contains the following six parts:
\begin{itemize}
	\item \texttt{PART~1: MATERIAL AND FLOW PARAMETERS}
	\item \texttt{PART~2: FINITE ELEMENT ANALYSIS PREPARATION and NON-DESIGN DOMAIN}
	\item \texttt{PART~3: PRESSURE \& STRUCTURE B.C's, LOADs}
	\item \texttt{PART~4: FILTER PREPARATION}
	\item \texttt{PART~5: MMA OPTIMIZATION PREPARATION}
	\item \texttt{PART~6: MMA OPTIMIZATION LOOP}
\end{itemize}
which are described in detail below.

\underline{\texttt{PART~1}} records the material and flow parameters \textbf{(lines~2-7)}.
$\mathtt{E1}$ is Young's modulus $E_1$ of the material, $\mathtt{Emin}$ is the Young's modulus assigned to the void elements to avoid numerical issues, and the Poisson's ratio is indicated via \texttt{nu}. \texttt{Kv} is the flow coefficient of void element, $K_v$, \texttt{epsf} is the flow contrast $\epsilon$, \texttt{Dels} is the penetration depth $\Delta s$, \texttt{Ds} is the drainage parameter $D_s$, and \texttt{kvs} is $K_v(1-\epsilon)$. Note, $\texttt{Kv}$ is set to 1, therefore ${K_s}=\epsilon$, i.e., $K_s = \texttt{epsf}$.
\begin{figure*}[h]
	\begin{subfigure}{0.5\textwidth}
		\centering
		\includegraphics[scale=0.85]{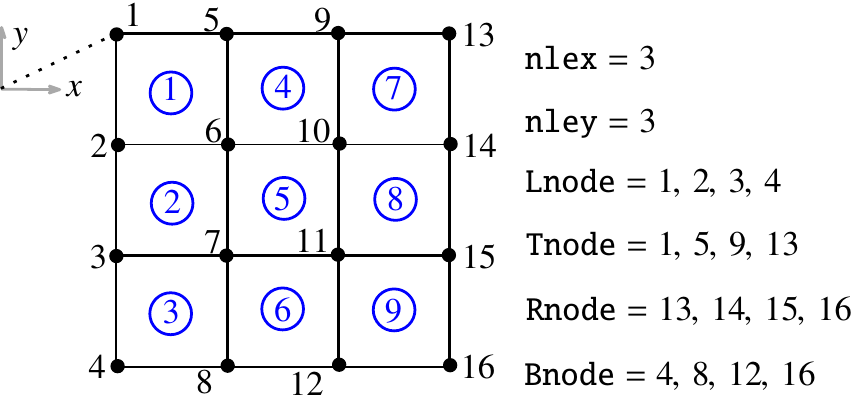}
		\caption{Mesh grid nomenclature: Global format}
		\label{fig:Gridinformation}
	\end{subfigure}
	\quad \quad
	\begin{subfigure}{0.5\textwidth}
		\centering
		\includegraphics[scale=0.85]{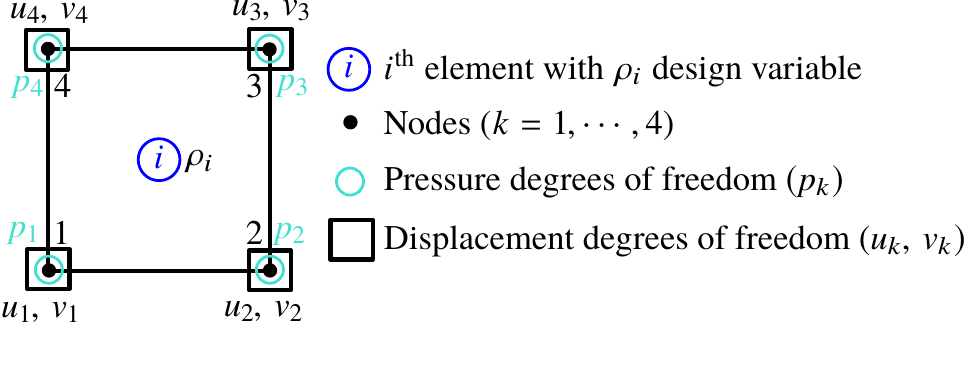}
		\caption{Nomenclature for element~$i$: Local format}
		\label{fig:nomenclatureSigle}
	\end{subfigure}
	\caption{Nomenclature for element~$i$ and a mesh grid are displayed. Each node is characterized by two displacements and one pressure degree of freedom. \texttt{nelx} and \texttt{nely} are the number of elements used in $x-$ and $y-$directions, respectively. \texttt{Lnode}, \texttt{Tnode}, \texttt{Rnode} and \texttt{Bnode} indicate  arrays of nodes constituting the left, top, right and bottom edges, respectively.} \label{fig:Nomandgridinformation}
\end{figure*}

\texttt{\underline{PART~2}} provides finite element analysis preparation steps \textbf{(lines~8-33)}.
Figure~\ref{fig:Gridinformation} displays terminology and style for nodes and elements numbering employed for the discretized domains\footnote{We use $\texttt{nelx}\times \texttt{nely} = 3 \times 3$ in Fig.~\ref{fig:Gridinformation} for the sake of visibility}. The local nomenclature and numbering for element~$i$ are depicted in Fig.~\ref{fig:nomenclatureSigle}. $\rho_i$ is the design variable that is considered constant within element~$i$ and thus, its physical density $\tilde{{\rho_i}}$ (Sec.~\ref{Sec:TopOptFor}).  Each element poses four nodes and two and one degrees of freedom (DOFs) per node for the displacement and pressure fields, respectively (Fig.~\ref{fig:nomenclatureSigle}). Node~$k$ poses $2k-1$ and $2k$ DOFs for the displacement in $x-$ and $y-$directions, respectively. The node number indicates the respective pressure DOF. The total number of elements and nodes are evaluated on line~9, which are denoted via \texttt{nel} and \texttt{nno}, respectively. Lines~10-12 are as per~\cite{andreassen2011efficient} where matrix \texttt{Udof} contains element-wise displacement DOFs. Row~$i$ of \texttt{Udof} gives the DOFs of nodes of element~$i$.

Nodes constituting the left, top, right, and bottom edges of the domains are recorded in arrays \texttt{Lnode}, \texttt{Tnode}, \texttt{Rnode} and \texttt{Bnode}, respectively (lines~13-14). Element-wise pressure DOFs are recorded in~\texttt{Pdofs} (line~15). Arrays \texttt{allPdofs} and \texttt{allUdofs} record pressure and displacement DOFs of the parameterized domains. Vectors \texttt{iP} and \texttt{jP} record the rows and columns indices for the flow matrix using \texttt{Pdofs} and \texttt{nel} (lines~16-17). Likewise, $\left\{\texttt{iT,\,jT}\right\}$ and $\left\{\texttt{iK,\,jK}\right\}$ contain the rows and columns indices for the transformation and stiffness matrices, respectively (lines~18-21). These vectors are defined per~\cite{andreassen2011efficient} as described for $\left\{\texttt{iK,\,jK}\right\}$.

Next, \texttt{Kp} determines the element Darcy flow matrix $\mathbf{K}_\text{p}^e$~(\ref{Eq:FEAnumint}) with unit flow coefficient on line~22. Likewise, \texttt{KDp} evaluates the element drainage matrix $\mathbf{K}_\text{DP}^e$~(\ref{Eq:FEAnumint}) with unit $D_s$ on line~23. Lines~25-29 provide a routine to determine the element stiffness matrix with unit Young's modulus on line~29 per~\cite{andreassen2011efficient}. Lines~30-33 provide a function handle for the Heaviside function and its derivative. Line~34 records the elements constituting non-design solid  and non-design void  regions in \texttt{NDS} and \texttt{NDV}, respectively. \texttt{act} determined on line~35 contains the active set of elements.

\texttt{\underline{PART~3}} describes the boundary conditions for the flow and structure settings \textbf{(lines~36-44)}. \texttt{PF} and \texttt{Pin} contain the initialize pressure vector and input pressure load (line~37). Line~38 applies the given pressure loading conditions for the arc problem. One needs to modify this line as per the different problems. The fixed pressure DOFs and the given corresponding pressure values are recorded in vectors \texttt{fixedPdofs} and \texttt{pfixeddofsv}, respectively. Vector \texttt{fixedUdofs} (line~42) contains the displacement DOFs, and corresponding free DOFs are determined on line~43 (vector \texttt{freeUdofs}). Line~44 initializes the displacement vector $\mathbf{u}$ and the Lagrange multiplier $\bm{\lambda}_1$ (\ref{Eq:OPTI}). 

\underline{\texttt{PART~4}} provides filter preparation per~\cite{ferrari2021topology} using the in-built \texttt{imfilter} function \textbf{(lines~45-48)}. The function allows to specify zero-Dirichlet or zero-Neumann boundary conditions. Here, we use the former, the default option for \texttt{imfilter}. 

\underline{\texttt{PART~5}} provides MMA optimization preparation, initialization, and allocations for some variables \textbf{(lines~49-58)}. \texttt{x} (line~50) and \texttt{xphys} (line~52) denote the actual and physical design variables, respectively. Vector \texttt{x} is first initialized to a zero vector (line~50). Then it is updated per the non-design solid and void regions (line~51), i.e., for the active set of elements.  Variables mentioned on lines~52-57 are specifically related to the MMA optimizer~\citep{svanberg1987}. Scalars \texttt{nMMA}  and \texttt{mMMA} indicate the number of active design variables and constraints, respectively (line~52). The design variable vector for the MMA is denoted by \texttt{xMMA}, which is initialized to the active design variable vector, $\texttt{x(act)}$. \texttt{xphys} is initialized on line~52. \texttt{mvLt} (line~52) is the external move limit for the MMA optimizer. On line~53, the minimum and maximum values of the design vector are defined by \texttt{xminvec} and \texttt{xmaxvec}, respectively.  Vectors \texttt{low} and \texttt{upp} contain the lower and upper limits of the design variables, respectively (line~54). The MMA constants, \texttt{cMMA}, \texttt{dMMA}, \texttt{a0}  and \texttt{aMMA} are defined on line~55. Vectors \texttt{xold1} and \texttt{xold2} contain the old value of the design vector \texttt{xMMA}, which are initialized on line~56. The optimization loop counter is recorded in \texttt{loop} (line~57), and the absolute change in the design vector is read in \texttt{change} (line~57). The derivative of the volume constraint \texttt{dvol0} is determined on line~50 and filtered on line~58, which is recorded in \texttt{dvol}.

\underline{\texttt{PART~6}} is executed on \textbf{lines~59-100} and contains five sub-parts. All variables within this part are repeatedly determined within the optimization loop. \texttt{sparse} function is employed for assembling the element flow, transformation and stiffness matrices using the vectors $\left\{\texttt{iP,\,jP}\right\}$, $\left\{\texttt{iT,\,jT}\right\}$, and $\left\{\texttt{iK,\,jK}\right\}$, respectively.

\underline{\texttt{Subpart~6.1}} solves the flow balance equation (\ref{Eq:PDEsolutionpressure}) on \textbf{lines~62-69}.  Using the Heaviside function handle (line~30), the flow (\texttt{Kc}) and drainage (\texttt{Dc}) coefficients of elements are determined on line~63 and line~64, respectively. \texttt{Kc} and \texttt{Dc} indicate $K$ and $D$ (\ref{Eq:stateequation}), respectively. The design-dependent flow vector is determined on line~65. \texttt{Kp} and \texttt{KDp} are converted into column vectors, are multiplied respectively by \texttt{Kc} (flow coefficient K~(\ref{Eq:Flowcoefficient})) and \texttt{Dc} (Drainage term D~(\ref{Eq:stateequation})) and then added to evaluate flow vector \texttt{Ae}. The global flow matrix \texttt{AG} (used to represent $\mathbf{A}$~(\ref{Eq:PDEsolutionpressure})) is determined on line~66 by placing \texttt{Ae} vector as third entries. Line~66 determines the flow matrix \texttt{Aff} that corresponds to free pressure DOFs. \texttt{PF}, i.e., $\mathbf{p}$~(\ref{Eq:PDEsolutionpressure}) is evaluated on line~67 using \texttt{decomposition} MATLAB function. On line~69, the vector \texttt{PF} is modified per the given pressure boundary conditions. 

\texttt{\underline{Subpart~6.2}} solves the structure balance equation (\textbf{lines~70-77}). The first part determines the global transformation matrix $\mathbf{T}$~(\ref{Eq:nodalforce}) denoted via \texttt{TG} (line~72) from the vector \texttt{Ts}. \texttt{Ts} is evaluated by converting $\mathbf{T}_e$~(\ref{Eq:Forcepressureconversion}) indicated by \texttt{Te} matrix into a column vector and appropriately reshaping it (line~71). The global consistent force vector is determined on line~73. Design-dependent Young's modulus vector \texttt{E} is determined on line~74. The stiffness vector \texttt{Ks} (line~75) is evaluated as \texttt{Ae} (line~65) using  element stiffness matrix~\texttt{ke} and \texttt{E}. The global stiffness matrix $\mathbf{K}$ indicated via \texttt{KG} is determined on line~76. The global displacement vector $\mathbf{u}$ denoted by \texttt{U} is determined on line~77 using the \texttt{decomposition} function with Cholesky.

\underline{\texttt{Subpart~6.3}} determines the objective, its sensitivities, and volume constraint (\textbf{lines~78-87}). Compliance~\eqref{Eq:OPTI} of the domain is minimized that is evaluated on line~79. Vector \texttt{lam1} indicate the Lagrange multiplier $\bm{\lambda}_1$ \eqref{Eq:LagrangeMultiplier} which is determined on line~80. Note that $\bm{\lambda}_2$ equals to -$2\mathbf{u}$. Vector \texttt{objsT1} contains the first term of the right-hand side of \eqref{Eq:senstivities_Obj}. Note that line~82 and line~83 evaluate the sensitivity terms pertaining to the Darcy flow and drainage term, respectively. These are added to determine $\texttt{objsT1}$. Likewise, the load sensitivities are evaluated and recorded in vector \texttt{objsT2} (line~84). Vector \texttt{objsens} (line~84) contains the total objective sensitivities (\ref{Eq:senstivities_Obj}). The volume constraint is determined on line~85 (scalar \texttt{Vol}). We normalize the objective and hence its sensitivities. The normalization scalar \texttt{normf} is determined on line~86 using the initial (\texttt{loop}=1) objective value. Objective sensitivities are filtered with normalization on line~87. However, one can prefer to perform normalization after evaluating the filtered sensitivities.
\begin{figure*}[ht!]
	\begin{subfigure}{0.45\textwidth}
		\centering
		\includegraphics[scale=0.65]{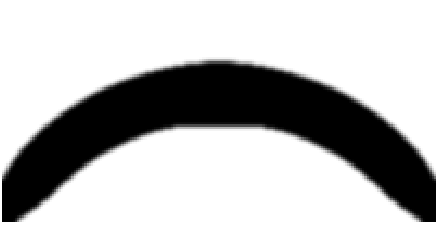}
		\caption{$C=36.25$}
		\label{fig:arch_mat}
	\end{subfigure}
	\quad
	\begin{subfigure}{0.45\textwidth}
		\centering
		\includegraphics[scale=0.65]{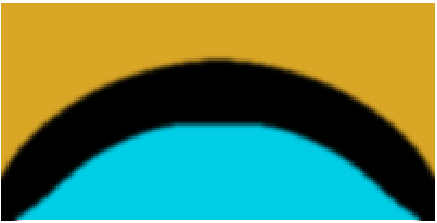}
		\caption{}
		\label{fig:arch_mat_pres}
	\end{subfigure}
	\caption{Optimized designs for internally pressurized arch structure with \texttt{lst}=1. (\subref{fig:arch_mat}) Optimized arch design and (\subref{fig:arch_mat_pres}) Optimized material layout with the final pressure field} \label{fig:Arch_Internally}
\end{figure*}

\underline{\texttt{Subpart~6.4}}  executes the optimization and updates of the design variables (\textbf{lines~88-96}). The \texttt{mmasub} (line~91) function is called wherein \texttt{xval} is initialized to \texttt{xMMA}. Vectors \texttt{xminvec} and \texttt{xmaxvec} are updated on line~90 using \texttt{mvLt}, the external move limit (line~52). On line~92, vectors \texttt{xold2}, \texttt{xold1} and \texttt{xnew} are updated using the \texttt{deal} MATLAB function. Line~95 updates the vectors \texttt{xMMA} and \texttt{xphys} using the new solution \texttt{xnew} provided by the MMA, and determines the filtered design variables. On line~96, \texttt{xphys} is updated with respect to the non-design solid and void elements. 

We use \texttt{MMA} (\texttt{mmasub}) MATLAB code written in 1999 and updated in 2002 by~\cite{svanberg1987}. The optimized structures may remain the same; however, their final objective values may differ (a little) due to numerical calculations and different parameter values used inside the MMA code if one uses a different version of the MMA. One can also choose to define~\texttt{(PART~4)} and apply (lines~58,\,87,\,95) the density filtering in a classical manner per~\cite{sigmund200199,andreassen2011efficient} and thereof, may get (slightly) different final objective values for the optimized designs due to numerical calculations and approximations. 

\underline{\texttt{subpart~6.5}} prints and plots the results using the \texttt{fprintf} and \texttt{imagesc} (\textbf{lines~97-99}). To print the material distribution with the pressure field, one can place the following lines of code below the end of \texttt{TOPress}:

\begin{lstlisting}[basicstyle=\scriptsize\ttfamily,breaklines=true,numbers=none,frame=tb]
	[xx,yy,nod_dV] = deal(0:nelx, 0:-1:-nely,zeros(nno,1));
	convert =[1; 1/2*ones(nely-1,1);1;reshape([1/2*ones(nelx-1,1),1/4*ones(nelx-1,nely-1)...
	,1/2*ones(nelx-1,1)]',[],1);1; 1/2*ones(nely-1,1); 1];
	PFP = figure(2); set(PFP,'color','w');
	for i = 1:nel, nod_dV(Pdofs(i,:)) = nod_dV(Pdofs(i,:)) + xphys(i); end
	xphysnodal = full(reshape(nod_dV.*convert,nely+1,nelx+1)); % nodal density grid
	PnodMat   = -full(reshape((PF)/Pin, nely+1, nelx+1));% nodal pressure field
	winter1 = [ linspace(0,0.85)' linspace(0.8078,0.65)' linspace(0.9,0.15)']; % colorbar style
	colormap([winter1; 0. 0. 0. ]); black = 0.01* ones(nely+1, nelx+1);
	PrDenHand = surf(xx,yy,PnodMat, 'FaceColor', 'interp'); caxis([-1 0.0]);hold on
	Hbl = imagesc(xx, yy, black, [-1 0.01]);  shading interp;xlim ([0 nelx]); ylim([-nely, 0]); view(0,90);
	PwDenHand = Hbl; axis equal off; set(PwDenHand, 'AlphaData', xphysnodal); hold on;
	set(PrDenHand, 'CData', PnodMat); set(PrDenHand,'ZData',PnodMat);
\end{lstlisting}
Next, we solve the internally pressurized arch structure by \texttt{TOPress} and then, the 100-line code is modified for different problems.

\subsection{The internally pressurized arch design}\label{Sec:Interpressarch}
To get the optimized designs for the loadbearing arch design (Fig.~\ref{fig:2Darcproblem}), we call \texttt{TOPress}  as
\begin{lstlisting}[basicstyle=\scriptsize\ttfamily,breaklines=true,numbers=none,frame=tb]
	TOPress(nelx,nely,volf,penal,rmin,etaf,betaf,lst,maxit)
\end{lstlisting}
with the following data. $\texttt{nelx} = 200$, \texttt{nely}=100, $\texttt{volf} =0.30$, $\texttt{penal}=3$, $\texttt{rmin} = 2.4$, $\texttt{etaf} = 0.2$, $\texttt{betaf}= 8$, $\texttt{lst}=1$ (load sensitivities are included), $\texttt{maxit} =100$ and \texttt{change}$>0.01$ are set. 

\begin{figure}[ht!]
	\centering
	\begin{tikzpicture}
		\pgfplotsset{compat = 1.3}
		\begin{axis}[ blue,
			width = 0.60\textwidth,
			xlabel=MMA iteration,
			axis y line* = left,
			ylabel= Compliance,
			ymajorgrids=true,
			xmajorgrids=true,
			grid style=dashed ]
			\pgfplotstableread{arch_volf_30_ls_1.txt}\mydata;
			\addplot[smooth,blue,mark = *,mark size=1pt,style={thick}]
			table {\mydata};\label{plot1}
		\end{axis}
		\begin{axis}[
			width = 0.60\textwidth,
			axis y line* = right,
			ylabel= Volume fraction,
			axis x line*= top,
			xlabel=MMA iteration,
			ytick = {0.290,0.295,0.300},
			yticklabel style={/pgf/number format/.cd,fixed,precision=3},
			legend style={at={(0.95,0.5)},anchor=east}]
			\addlegendimage{/pgfplots/refstyle=plot1}\addlegendentry{Objective}
			\pgfplotstableread{arch_volf_30_ls_1_VF.txt}\mydata;
			\addplot[smooth,black,mark = square,mark size=1pt,style={thick}]
			table {\mydata};
			\addlegendentry{volume fraction}
		\end{axis}
	\end{tikzpicture}
	\caption{Convergence plot for the internally pressurized arch problem}
	\label{fig:ConvergenceVolf30}
\end{figure}
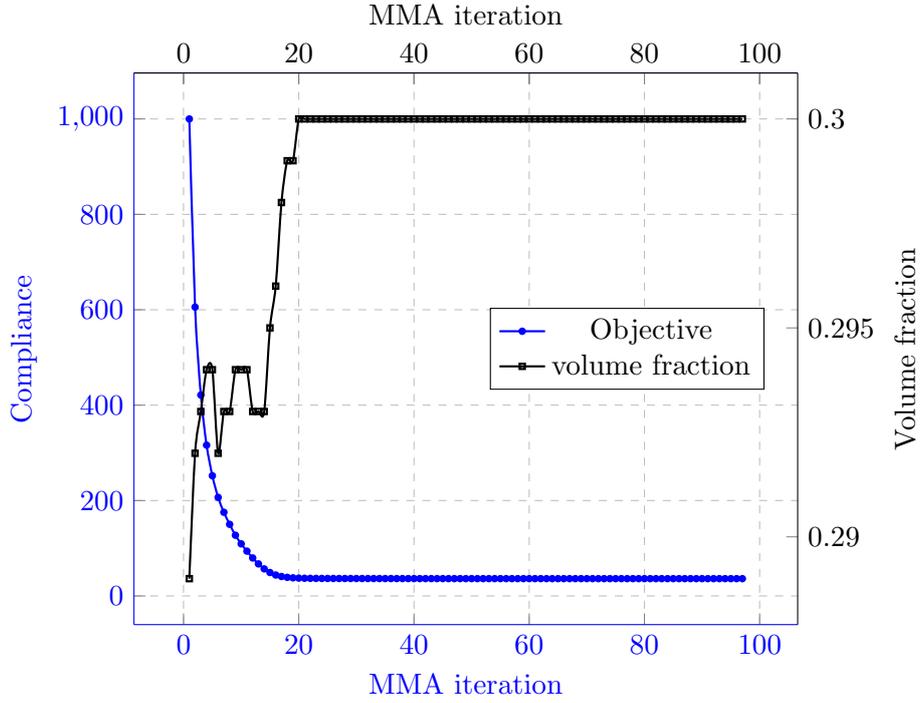

The optimized arch design, the final P-field with the optimum material layout and convergence curve are depicted in Fig.~\ref{fig:arch_mat}, Fig.~\ref{fig:arch_mat_pres} and Fig.~\ref{fig:ConvergenceVolf30}, respectively. The arch problem is converged at 97$^\text{th}$ MMA iteration. Note that for plotting the final pressure field with the optimized material layout, the code mentioned above is inserted at the end of the 100-line code. The objective convergence is smooth and rapid (Fig.~\ref{fig:ConvergenceVolf30}). The volume constraint is satisfied and remains active at the end of the optimization (Fig.~\ref{fig:ConvergenceVolf30}).

Next, we present a study with different volume fractions with and without load sensitivity terms to understand their effects on the optimized designs. We use four volume fractions, \texttt{volf} =0.05,\,0.1,\,0.25,\, and 0.5. $\left\{\texttt{etaf,\,betaf}\right\}$ are set to $\left\{\texttt{volf},\,6\right\}$~\citep{kumar2020topology}. Other parameters are taken the same as employed above.

\begin{figure*}[h!]
	\begin{subfigure}{0.22\textwidth}
		\centering
		\texttt{volf=0.05}
		\begin{tikzpicture}
			\node[anchor=south west,inner sep=0] at
			(0,0){\includegraphics[scale=0.35]{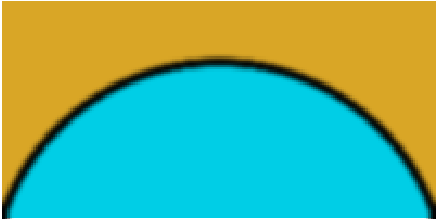}};
			\draw[blue,line width = 0.75pt] (1.6,1.53)--(5,1.53);
		\end{tikzpicture}
		\caption*{$C=1.12,\,\texttt{lst}=1$}
		\label{fig:Arch_volf_05_ls_1_mat_pre}
	\end{subfigure}
	\quad
	\begin{subfigure}{0.22\textwidth}
		\centering
		\texttt{volf=0.05}
		\begin{tikzpicture}
			\node[anchor=south west,inner sep=0] at
			(0,0){\includegraphics[scale=0.35]{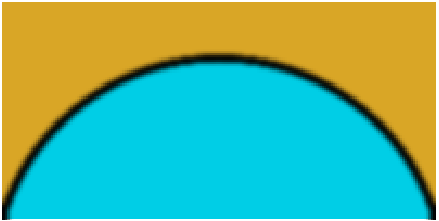}};
			\draw[red,dashed,line width = 0.75pt] (0,1.60)--(2.5,1.60);
		\end{tikzpicture}
		\caption*{$C=1.14,\,\texttt{lst}=0$}
		\label{fig:Arch_volf_05_ls_0_mat_pre}
	\end{subfigure}
	\quad
	\begin{subfigure}{0.22\textwidth}
		\centering
		\texttt{volf=0.1}
		\begin{tikzpicture}
			\node[anchor=south west,inner sep=0] at
			(0,0){\includegraphics[scale=0.35]{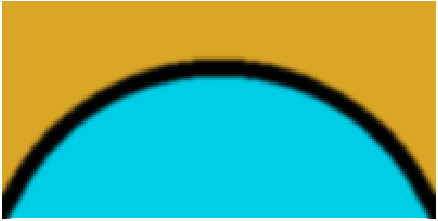}};
			\draw[blue,line width = 0.75pt] (1.6,1.50)--(5,1.50);
		\end{tikzpicture}
		\caption*{$C=3.43,\,\texttt{lst}=1$}
		\label{fig:Arch_volf_1_ls_1_mat_pre}
	\end{subfigure}
	\quad
	\begin{subfigure}{0.22\textwidth}
		\centering
		\texttt{volf=0.1}
		\begin{tikzpicture}
			\node[anchor=south west,inner sep=0] at
			(0,0){\includegraphics[scale=0.35]{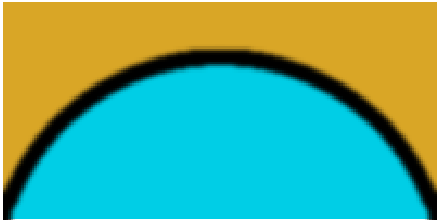}};
			\draw[red,dashed,line width = 0.75pt] (0,1.63)--(2.5,1.63);
		\end{tikzpicture}
		\caption*{$C=3.57,\,\texttt{lst}=0$}
		\label{fig:Arch_volf_1_ls_0_mat_pre}
	\end{subfigure}

	\begin{subfigure}{0.22\textwidth}
		\vspace{1.5em}
		\centering
		\texttt{volf=0.25}
		\begin{tikzpicture}
			\node[anchor=south west,inner sep=0] at
			(0,0){\includegraphics[scale=0.35]{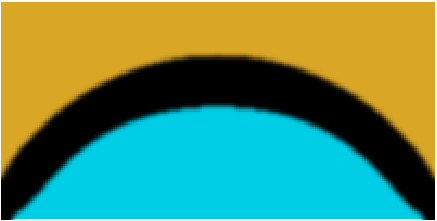}};
			\draw[blue,line width = 0.75pt] (1.6,1.55)--(5,1.55);
		\end{tikzpicture}
		\caption*{$C=23.58,\,\texttt{lst}=1$}
		\label{fig:Arch_volf_25_ls_1_mat_pre}
	\end{subfigure}
	\quad\,
	\begin{subfigure}{0.22\textwidth}
		\vspace{1.5em}
		\centering
		\texttt{volf=0.25}
		\begin{tikzpicture}
			\node[anchor=south west,inner sep=0] at
			(0,0){\includegraphics[scale=0.35]{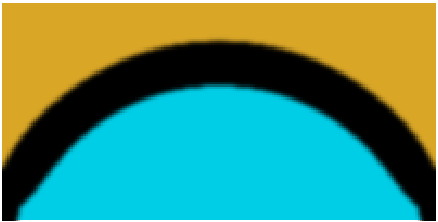}};
			\draw[red,dashed,line width = 0.75pt] (-0.1,1.71)--(2.5,1.71);
		\end{tikzpicture}
		\caption*{$C=24.62,\,\texttt{lst}=0$}
		\label{fig:Arch_volf_25_ls_0_mat_pre}
	\end{subfigure}
	\quad 
	\begin{subfigure}{0.22\textwidth}
		\vspace{1.5em}
		\centering
		\texttt{volf=0.5}
		\begin{tikzpicture}
			\node[anchor=south west,inner sep=0] at
			(0,0){\includegraphics[scale=0.35]{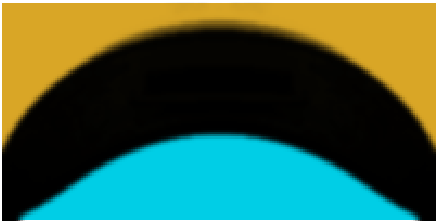}};
			\draw[blue,line width = 0.75pt] (1.6,1.86)--(5,1.86);
		\end{tikzpicture}
		\caption*{$C=135.14,\,\texttt{lst}=1$}
		\label{fig:Arch_volf_5_ls_1_mat_pre}
	\end{subfigure}
	\quad
	\begin{subfigure}{0.22\textwidth}
		\vspace{1.5em}
		\centering
		\texttt{volf=0.5}
		\begin{tikzpicture}
			\node[anchor=south west,inner sep=0] at
			(0,0){\includegraphics[scale=0.35]{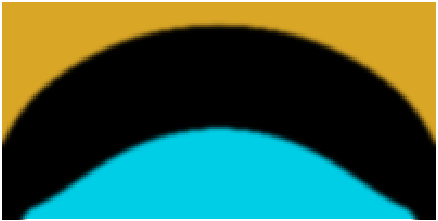}};
			\draw[red,dashed,line width = 0.75pt] (0,1.87)--(2.5,1.87);
		\end{tikzpicture}
		\caption*{$C=135.80,\,\texttt{lst}=0$}
		\label{fig:Arch_volf_5_ls_0_mat_pre}
	\end{subfigure}
	\caption{Optimized designs for internally pressurized arch structures for various volume fractions with \texttt{lst}=1 and \texttt{lst} =0.} \label{fig:Arch_Internally2}
\end{figure*}
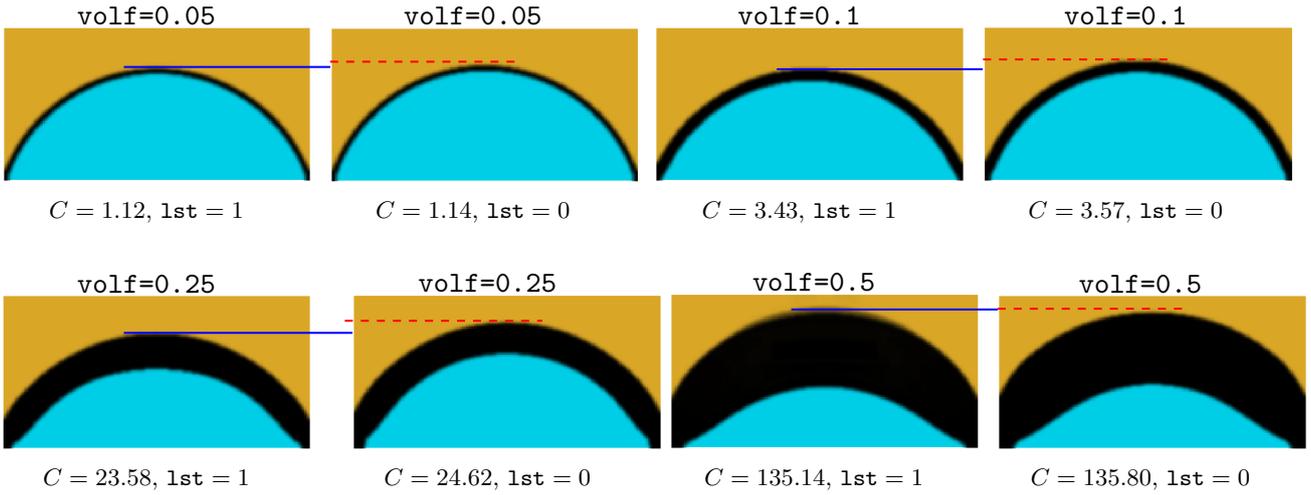

\begin{figure}[h]
	\begin{subfigure}{0.3\textwidth}
		\centering
		\includegraphics[scale=0.89]{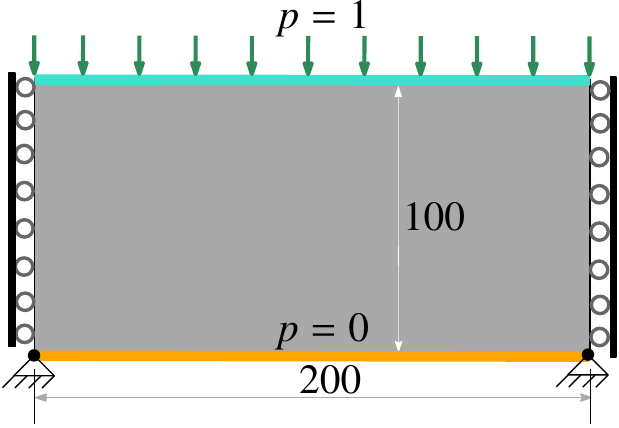}
		\caption{}
		\label{fig:Bridge_structure}
	\end{subfigure}
	\quad \quad
	\begin{subfigure}{0.30\textwidth}
		\begin{tikzpicture}
			\node[anchor=south west,inner sep=0] at
			(0,0){\includegraphics[scale=0.44]{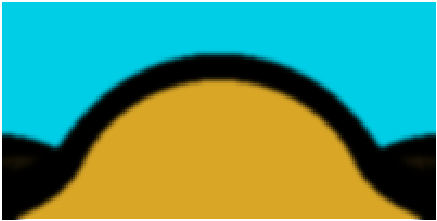}};
			\draw[blue,line width = 1pt] (2.5,1.97)--(5.5,1.97);
		\end{tikzpicture}
		\caption{$C=14.48,\,\texttt{lst}=1$}
		\label{fig:Bridge_mat_Pre_ls_1}
	\end{subfigure}
	\quad
	\begin{subfigure}{0.30\textwidth}
		\begin{tikzpicture}
			\node[anchor=south west,inner sep=0] at
			(0,0){\includegraphics[scale=0.44]{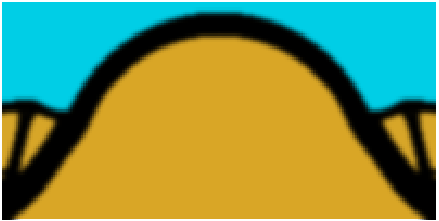}};
			\draw[red,dashed,line width = 1pt] (-0.25,2.45)--(3,2.45);
		\end{tikzpicture}
		\caption{$C=15.30,\,\texttt{lst}=0$}
		\label{fig:Bridge_mat_Pre_ls_0}
	\end{subfigure}
	\caption{The design domain, pressure and displacement boundary conditions for the bridge-like structure is depicted in~(\subref{fig:Bridge_structure}). Optimized results with \texttt{lst}=1 and \texttt{lst}=0 are displayed in (\subref{fig:Bridge_mat_Pre_ls_1}) and (\subref{fig:Bridge_mat_Pre_ls_0}), respectively.}  \label{fig:Bridge_strucutreall}
\end{figure}

The optimized designs of the arch structure with (\texttt{lst}=1) and without (\texttt{lst}=0) load sensitivities are displayed in  Fig.~\ref{fig:Arch_Internally2}. One can note that optimized designs obtained with \texttt{lst}=1 are relatively stiffer, i.e., high performing (lowered in height) than their counterparts obtained with \texttt{lst=0}. Though the final topologies with \texttt{lst=1} and with \texttt{lst}=0 are, by and large,  same for this problem, and their performances are also close to each other, neglecting load sensitivities~\eqref{Eq:senstivities_Obj} may not be a good idea. This gets readily confirmed by other forthcoming numerical examples presented below. The difference in the height of the arch designs is directly related to the final compliance difference for \texttt{lst}=0 and \texttt{lst}=1.  \texttt{TOPress} provides good results even for relatively low volume fraction (Fig.~\ref{fig:Arch_Internally2}). One can note that $C$ increases with the volume fraction, which should be the case. Next, we modify \texttt{TOPress} as per the pressure and displacement boundary conditions of different problems of interest.

\subsection{Bridge-like pressure loadbearing structure}
Figure~\ref{fig:Bridge_structure} displays the design domain, pressure, and displacement boundary conditions for a bridge-like loadbearing structure per~\cite{du2004topological}. Pressure load is applied on the top edge, whereas the bottom edge is assigned zero pressure load. Left and right edges are permitted to move in the vertical direction. For this problem, one needs to make the following modifications in \texttt{TOPress}. Line~38 is changed to
\begin{lstlisting}[basicstyle=\scriptsize\ttfamily,breaklines=true,numbers=none,frame=tb]
	PF(Bnode) = 0; PF(Tnode) = Pin; % applying pressure load
\end{lstlisting}
and line~42 is altered to
\begin{lstlisting}[basicstyle=\scriptsize\ttfamily,breaklines=true,numbers=none,frame=tb]
	fixedUdofs = [2*Bnode(1)-1  2*Bnode(1)  2*Bnode(end)-1 2*Bnode(end) 2*Lnode-1 2*Rnode-1]; %fixed displ.
\end{lstlisting}
With the above changes, the optimized design depicted in Fig.~\ref{fig:Bridge_mat_Pre_ls_1} and Fig.~\ref{fig:Bridge_mat_Pre_ls_0} respectively are obtained by the following call:
\begin{lstlisting}[basicstyle=\scriptsize\ttfamily,breaklines=true,numbers=none,frame=tb]
	TOPress(200,100,0.20,3,2.5,0.20,10,1,150) 	
\end{lstlisting}
and
\begin{lstlisting}[basicstyle=\scriptsize\ttfamily,breaklines=true,numbers=none,frame=tb]
	TOPress(200,100,0.20,3,2.5,0.20,10,0,150).
\end{lstlisting}
The structure obtained with \texttt{lst}=1 (Fig.~\ref{fig:Bridge_mat_Pre_ls_1}) is stiffer (high performing) and has less arch height than that obtained with \texttt{lst}=0 (Fig.~\ref{fig:Bridge_mat_Pre_ls_0}). The observation is similar to what we find for internally pressurized arch structures. In addition, they have different topologies (Fig.~\ref{fig:Bridge_mat_Pre_ls_1} and Fig.~\ref{fig:Bridge_mat_Pre_ls_0}). Thus, load sensitivities affect the final topology and can help improve the performance of the optimized designs.It may happen that for certain value of $\left\{\eta _f,\,\beta _f\right\}$, one may get optimized designs with leakage. Therefore, one may follow the recommendation given for $\left\{\eta_f,\,\beta_f\right\}$ in~\cite{kumar2020topology} to achieve a leak-proof optimized design.
\begin{figure}[h]
	\begin{subfigure}{0.3\textwidth}
		\centering
		\includegraphics[scale=0.870]{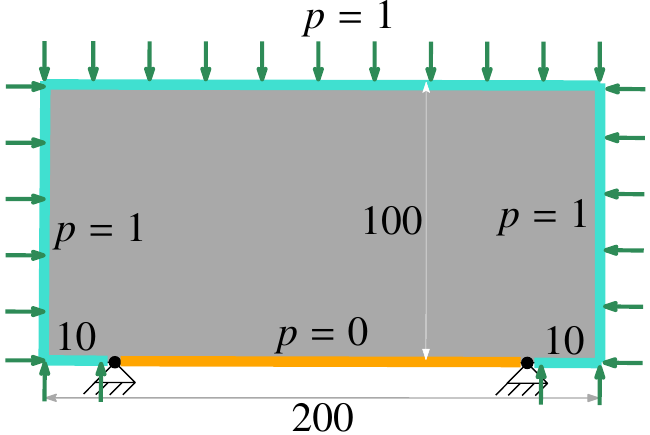}
		\caption{}
		\label{fig:Arch_ext_structure}
	\end{subfigure}
	\quad \quad
	\begin{subfigure}{0.30\textwidth}
		\begin{tikzpicture}
			\node[anchor=south west,inner sep=0] at
			(0,0){\includegraphics[scale=0.45]{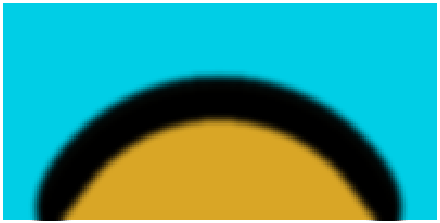}};
			\draw[blue,line width = 1pt] (2.5,1.8)--(5.5,1.8);
		\end{tikzpicture}
		\caption{$C=17.80,\,\texttt{lst}=1$}
		\label{fig:arch_ext_mat_Pre_ls_1}
	\end{subfigure}
	\quad
	\begin{subfigure}{0.30\textwidth}
		\begin{tikzpicture}
			\node[anchor=south west,inner sep=0] at
			(0,0){\includegraphics[scale=0.45]{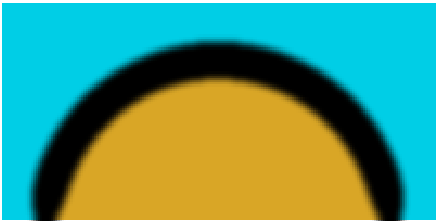}};
			\draw[red,dashed,line width = 1pt] (-0.25,2.22)--(3,2.22);
		\end{tikzpicture}
		\caption{$C=19.81,\,\texttt{lst}=0$}
		\label{fig:arch_ext_mat_Pre_ls_0}
	\end{subfigure}
	\caption{(\subref{fig:Arch_ext_structure}) The design domain, pressure and displacement boundary conditions for the bridge-like structure. In (\subref{fig:Bridge_mat_Pre_ls_1}) and (\subref{fig:Bridge_mat_Pre_ls_0}) show the optimized results with \texttt{lst}=1 and \texttt{lst}=0, respectively.}  \label{fig:Arch_externally_all}
\end{figure}

\begin{figure}[h]
	\begin{subfigure}{0.3\textwidth}
		\centering
		\includegraphics[scale=0.78]{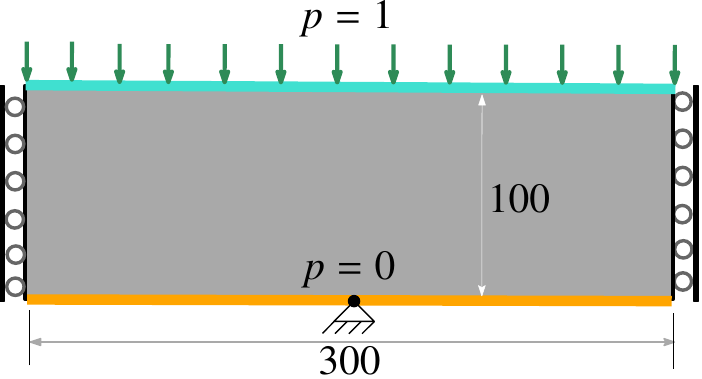}
		\caption{}
		\label{fig:piston_head_str}
	\end{subfigure}
	\quad \quad
	\begin{subfigure}{0.3\textwidth}
		\centering
		\includegraphics[scale=0.45]{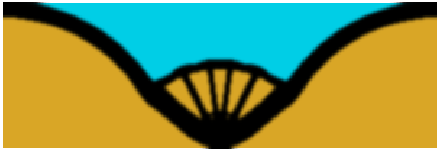}
		\caption{$C=17.72,\,\texttt{lst}=1$}
		\label{fig:piston_mat_Pre_ls_1}
	\end{subfigure}
	\quad 
	\begin{subfigure}{0.3\textwidth}
		\centering
		\includegraphics[scale=0.45]{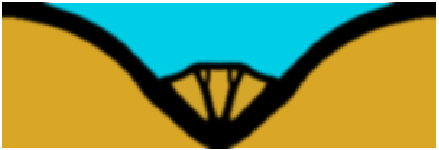}
		\caption{$C= 17.65,\,\texttt{lst}=0$}
		\label{fig:piston_mat_Pre_ls_0}
	\end{subfigure}
	\caption{(\subref{fig:piston_head_str}) The piston head the design domain with pressure and displacement boundary conditions. (\subref{fig:piston_mat_Pre_ls_1}) The optimized piston head design with \texttt{lst}=1, (\subref{fig:piston_mat_Pre_ls_0}) The optimized piston head design with \texttt{lst}=0}  \label{fig:piston_all}
\end{figure}
\subsection{Externally pressurized arch structure}
The externally pressurized arch problem was first solved in~\cite{hammer2000topology}, and subsequently it also appeared in~\cite{chen2001topology,du2004topological,sigmund2007topology, xia2015topology,emmendoerfer2018level,picelli2015bi,picelli2019topology,neofytou2020level}, to name a few. The design domain, pressure, and displacement boundary conditions for the problem are shown in Fig.~\ref{fig:Arch_ext_structure}. The left, right, top, and some parts' bottom edges are on pressure load $p=1$. Both ends of the bottom edge are fixed, as shown. The following lines of \texttt{TOPress} are modified to account for the boundary conditions associated with this problem. Line~38 (pressure input) is changed to
\begin{lstlisting}[basicstyle=\scriptsize\ttfamily,breaklines=true,numbers=none,frame=tb]
	PF([Tnode,Lnode,Rnode, Bnode(1:nelx/10), Bnode(end-nelx/10:end)]) = Pin; PF(Bnode(nelx/10+1:end-nelx/10-1)) = 0; % applying pressure 
\end{lstlisting}
and line~42 is modified to
\begin{lstlisting}[basicstyle=\scriptsize\ttfamily,breaklines=true,numbers=none,frame=tb]
	fixedUdofs = [2*Bnode(nelx/10+1)-1  2*Bnode(nelx/10+1)  2*Bnode(end-nelx/10-1)-1 2*Bnode(end-nelx/10-1)]; %fixed displ.
\end{lstlisting}

The optimized arch structures with \texttt{lst}=1 in Fig.~\ref{fig:arch_ext_mat_Pre_ls_1} and \texttt{lst}=0 in Fig.~\ref{fig:arch_ext_mat_Pre_ls_0} are  respectively obtained by the following function call
\begin{lstlisting}[basicstyle=\scriptsize\ttfamily,breaklines=true,numbers=none,frame=tb]
	TOPress(200,100,0.2,3,4,0.15,8,lst,100) 	
\end{lstlisting}
The optimized design with $\texttt{lst}=1$ has relatively lower-height arch and compliance (Fig.~\ref{fig:arch_ext_mat_Pre_ls_1}) than that obtained with $\texttt{lst}=0$ (Fig.~\ref{fig:arch_ext_mat_Pre_ls_0}). Including load sensitivities within the formulation may certainly affect the optimized shape and topology and thus, the final objective value.  The shape of the former resembles one presented in~\cite{sigmund2007topology}, whereas the shape of the latter resembles that reported in~\cite{picelli2019topology,neofytou2020level}.

\begin{figure}[h]
	\begin{subfigure}{0.3\textwidth}
		\centering
		\includegraphics[scale=0.80]{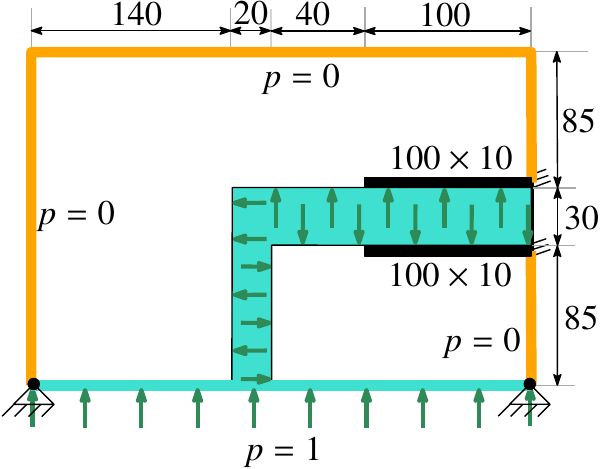}
		\caption{}
		\label{fig:Pressurized_chamber}
	\end{subfigure}
	\quad 
	\begin{subfigure}{0.3\textwidth}
		\centering
		\includegraphics[scale=0.40]{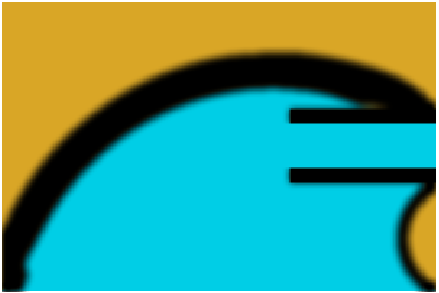}
		\caption{$C=11.78,\,\texttt{volf}=0.2$}
		\label{fig:pressurechamb_mat_pre_ls_1_volf_20}
	\end{subfigure}
	\quad 
	\begin{subfigure}{0.3\textwidth}
		\centering
		\includegraphics[scale=0.40]{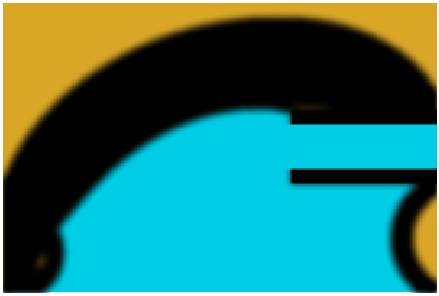}
		\caption{$C=68.52,\,\texttt{volf}=0.4$}
		\label{fig:pressurechamb_mat_pre_ls_1_volf_40}
	\end{subfigure}
	\caption{(\subref{fig:Pressurized_chamber}) The design domain for the pressurized chamber with pressure and displacement boundary conditions. (\subref{fig:pressurechamb_mat_pre_ls_1_volf_20}) and (\subref{fig:pressurechamb_mat_pre_ls_1_volf_40}) display the optimized optimized designs with \texttt{volf} $=0.2$ and \texttt{volf}$=0.4$, respectively.}  \label{fig:Pressurized_chamber_all}
\end{figure}

\subsection{Piston design}
We extend \texttt{TOPress} code for designing the pressure head here. This problem was first reported in~\cite{bourdin2003design}; after that, many papers also solved it, e.g., \cite{sigmund2007topology,picelli2015bi,emmendoerfer2018level,picelli2019topology,neofytou2020level,kumar2020topology,huang2022thermal}, to name a few. The design domain specification with pressure and displacement boundary conditions for the piston head model is shown in Fig.~\ref{fig:piston_head_str}. Pressure load is applied on the top edge, whereas the bottom edge is fixed and set at zero pressure load. The middle point of the bottom edge is fixed. The left and right edges are permitted to move in the $y-$axis. One makes the following changes in \texttt{TOPress} to solve this problem. Line~38 is changed to
\begin{lstlisting}[basicstyle=\scriptsize\ttfamily,breaklines=true,numbers=none,frame=tb]
	PF(Bnode) = 0; PF(Tnode) = Pin;  % applying pressure load	
\end{lstlisting}
and line~42 is modified to
\begin{lstlisting}[basicstyle=\scriptsize\ttfamily,breaklines=true,numbers=none,frame=tb]
	fixedUdofs = [2*Bnode(ceil((nelx+1)/2))-1  2*Bnode(ceil((nelx+1)/2))  2*Lnode-1 2*Rnode-1]; %fixed displ.	
\end{lstlisting}
The optimized designs depicted in Fig.~\ref{fig:piston_mat_Pre_ls_1}, \subref{fig:piston_mat_Pre_ls_0} that are obtained by the following call
\begin{lstlisting}[basicstyle=\scriptsize\ttfamily,breaklines=true,numbers=none,frame=tb]
	TOPress(300,100,0.2,3,2.4,0.1,8,lst,150)
\end{lstlisting}
with \texttt{lst=1} and \texttt{lst=0}, respectively. 

We consider the full model to solve this problem to check any deviation from the symmetric nature of the problem. One notes that the optimized design retains the symmetric nature (Fig.~\ref{fig:piston_mat_Pre_ls_1}) and qualitatively resembles 
the features of the piston structures reported in the previous literature. One may also solve this problem using only the symmetric half of the domain. For achieving leak-proof optimized pistons, the user may refer to \cite{kumar2020topology} for recommendations on choosing \texttt{etaf} and \texttt{betaf}. Different optimized material layouts are obtained with \texttt{lst}=1 and \texttt{lst}=0.  
\subsection{Pressurized chamber design} \label{Sec:Prechamdesign}
Herein \texttt{TOPress} is modified to solve the pressurized chamber design that was first presented by~\cite{hammer2000topology}. The problem is subsequently solved  in~\cite{chen2001topology,zhang2008new,picelli2015bi,picelli2019topology,neofytou2020level,huang2022thermal}. The design domain with pressure and displacement boundary conditions is shown in Fig.~\ref{fig:Pressurized_chamber}. The pressure load is applied using a chamber as indicated in the figure. The initial structure of the chamber is constructed through the non-design void and solid regions (Fig.~\ref{fig:Pressurized_chamber_all}). Both corners of the bottom edge and the right most edges of the solid non-design domains are fixed (Fig.~\ref{fig:Pressurized_chamber}). The following modifications one needs to perform to solve this problem with \texttt{TOPress}. Lines~34-35 are replaced by
\begin{lstlisting}[basicstyle=\scriptsize\ttfamily,breaklines=true,numbers=none,frame=tb]
	elNrs = reshape(1:nel,nely,nelx);
	sr1 = elNrs(3*nely/8:17*nely/40, 2*nelx/3:nelx);
	sr2 =  elNrs(23*nely/40:5*nely/8, 2*nelx/3:nelx);
	vr1 = elNrs(17*nely/40:end,7*nelx/15:8*nelx/15);
	vr2 = elNrs(17*nely/40:23*nely/40,8*nelx/15:nelx);
	[NDS, NDV] = deal( [sr1(:);sr2(:)], [vr1(:); vr2(:)] );      
	act = setdiff( (1 : nel)', union( NDS, NDV ) );
	s1fix = elNrs(3*nely/8:17*nely/40, nelx);
	s2fix = elNrs(23*nely/40:25*nely/40, nelx);
	fixx = unique(Pdofs([s1fix(:); s2fix(:)],:));	
\end{lstlisting}
where \texttt{elNrs} gives the matrix of elements for the parameterized domain. \texttt{sr1} and \texttt{sr2} provide elements in the solid regions (black bars, Fig.~\ref{fig:Pressurized_chamber}). Likewise, \texttt{vr1} and \texttt{vr2} record the elements forming void regions.  Pressurized regions are also depicted. \texttt{NDS} and \texttt{NDV} contain elements representing solid and void regions, respectively. The rightmost part of the solid regions contain elements recorded in vectors \texttt{s1fix} and \texttt{s2fix}. Vector \texttt{fixx} gives the nodes corresponding to \texttt{s1fix} and \texttt{s2fix}, which are fixed. Next, line~46 is changed to
\begin{lstlisting}[basicstyle=\scriptsize\ttfamily,breaklines=true,numbers=none,frame=tb]
	PF([Tnode,Lnode,Rnode]) = 0; PF([unique(Pdofs(NDV,:))',Bnode]) = Pin;% applying pressure load	
\end{lstlisting}
and line~50 is altered to
\begin{lstlisting}[basicstyle=\scriptsize\ttfamily,breaklines=true,numbers=none,frame=tb]
	fixedUdofs = [2*Bnode(1)-1  2*Bnode(1)  2*Bnode(end)-1 2*Bnode(end) 2*fixx'-1 2*fixx']; %fixed displ.
\end{lstlisting}
With the above modifications, the code is called as
\begin{lstlisting}[basicstyle=\scriptsize\ttfamily,breaklines=true,numbers=none,frame=tb]
	TOPress(300,200,volf,3,6,0.1,10,1,200),
\end{lstlisting}
The optimized results corresponding to \texttt{volf=0.2} and \texttt{volf =0.4} are shown in Fig.~\ref{fig:pressurechamb_mat_pre_ls_1_volf_20} and Fig.~\ref{fig:pressurechamb_mat_pre_ls_1_volf_40}, respectively. Features of the optimized designs resemble those reported in the previous research papers. The optimized shape of the pressure chamber changes into an arbitrary shape. In addition, the pressure chamber gets a small arch design at the rightmost part of the bottom edge, which is expected in accordance with the boundary conditions of the internally pressurized arch structure~(Fig.~\ref{fig:2Darcproblem}).  
\subsection{Heaviside projection filter}
The 100-line code is extended here with the Heaviside projection filter to achieve  the optimized pressure loadbearing designs close to 0-1~\cite{}. Three variables are used to represent each element: the actual variable ($\rho$), the filtered variable ($\tilde{\rho})$, and the projection variable ($\bar{\rho})$. The latter is determined for element~$i$ as
\begin{equation}\label{Eq:projectionFilt}
	\bar{\rho}_i = \frac{\tanh(\frac{\beta}{2}) + \tanh(\beta(\tilde{\rho}_i-\frac{1}{2}))} {2\tanh(\frac{\beta}{2})},
\end{equation}
where $\beta \in [0,\,\infty)$ is called the steepness parameter, and $\bar{\rho}_i$ is also termed the physical variable of element~$i$. We represent $\rho$, $\tilde{\rho}$ and $\bar{\rho}$ by $\mathtt{x}$,\, $\mathtt{xtilde}$ and $\mathtt{xPhys}$ in the extended code.  Here, $\beta$ is increased from $1$ to $\beta_\text{max}=256$ using a continuation scheme wherein $\beta$ gets doubled at every 25 MMA iterations. The chain rule is employed to determine the derivative of $\bar{\rho}$ with respect to $\rho$ and thus that of the objective and constraints. One determines derivatives of $\bar{\rho}$ with respect to $\tilde{\rho}$ as
\begin{equation}
	\frac{\partial \bar{\rho}_i}{\partial \tilde{{\rho_i}}} = \beta \frac{1 - \tanh (\beta(\tilde{{\rho_i}}-\frac{1}{2}))^2}{2\tanh(\frac{\beta}{2})}.
\end{equation}
\texttt{TOPress} is modified to cater to this filter for the internally pressurized arch design~(Sec.~3.1). One can adopt  similar steps for achieving black-and-white loadbearing designs for other problems with the projection filter. The following piece of code is inserted after line~57
\begin{lstlisting}[basicstyle=\scriptsize\ttfamily,breaklines=true,numbers=none,frame=tb]
	betap = 1;
	xPhys = IFprj(xTilde,0.5,betap);
\end{lstlisting}
where \texttt{betap} represents $\beta$ (\ref{Eq:projectionFilt}). Vector $\texttt{xTilde}$ that corresponds to the filtered variables is defined and initialized on line~52. The following line is inserted after line~62 to evaluate the filter matrix \texttt{dHs} as
\begin{lstlisting}[basicstyle=\scriptsize\ttfamily,breaklines=true,numbers=none,frame=tb]
	dHs = Hs./ reshape( dIFprj( xTilde, 0.5, betap ), nely, nelx);
\end{lstlisting}
As the matrix \texttt{dHs} varies in each iteration, the derivatives of volume constraint is now determined within the loop after line~86 as
\begin{lstlisting}[basicstyle=\scriptsize\ttfamily,breaklines=true,numbers=none,frame=tb]
	dVol = imfilter(reshape(dVol0, nely, nelx)./dHs,h); %filtered volume sensitivity
\end{lstlisting}
The objective sensitivities are accordingly determined  using \texttt{dHs}. Lines~96-97 are replaced by
\begin{lstlisting}[basicstyle=\scriptsize\ttfamily,breaklines=true,numbers=none,frame=tb]
	xMMA = xnew;   xTilde(act) = xnew; 
	xTilde = imfilter(reshape(xTilde, nely, nelx),h)./Hs;xTilde = xTilde(:);
	if(mod(loop,25)==0 && betap<=betamax), betap= betap*2;end
	xphys = IFprj(xTilde,0.5,betap);xphys(NDS)=1; xphys(NDV)=0;
\end{lstlisting}
wherein $\beta$ (\texttt{betap}) is doubled at each 25 MMA iterations until it reaches to $\beta_\text{max}$ (\texttt{betamax}).  The projection filtering and updation of $\beta$ ideally should be performed after line~99 as the employed projection filter is not volume preserving. However, to keep the printing/plotting position consistent and notwithstanding the small error introduced in printing values, we have included projection filtering on line~96-97. We now include $\beta_\text{max}$ (\texttt{betamax}) in the input of the main code
\begin{lstlisting}[basicstyle=\scriptsize\ttfamily,breaklines=true,numbers=none,frame=tb]
	TOPress(nelx,nely,volfrac,penal,rmin,etaf,betaf,betamax,lst,maxit)
\end{lstlisting}
One should set an appropriate value for $\beta_\text{max}$ so that it can be achieved within the assigned \texttt{maxit}. With the above modification, the code is called with $\texttt{volf}=0.3,\,\texttt{rmin}=4.8,\,\texttt{etaf} = 0.2,\,\texttt{betaf}=8,\,\texttt{betamax} = 256,\,\texttt{lst} =1$,  as
\begin{lstlisting}[basicstyle=\scriptsize\ttfamily,breaklines=true,numbers=none,frame=tb]
	TOPress(200,100,0.30,3,2.4,0.2,8,256,1,250)
\end{lstlisting}
to solve the internally pressurized arch design.
\begin{figure}[h!]
	\centering
	\begin{tikzpicture} 
		\node[inner sep=0pt] () at (-1,0){\includegraphics[scale=0.5]{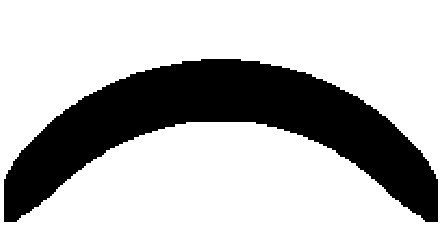} };
		\DrawBoundingBox[gray]
	\end{tikzpicture}
	\caption{Optimized internally pressurized arch structure with the Heaviside projection. $C =30.39$}\label{fig:archprojected}
\end{figure}

Fig.~\ref{fig:archprojected} displays the optimized material layout of the internally pressurized arch structure with the Heaviside projection filter.  The obtained gray scale parameter $M_{nd}$~\citep{sigmund2007} of the optimized structure is $\SI{8.77 e-6}{}$\% which indicates that the optimized arch structure is very close to binary.

\begin{figure}[h!]
	\centering
	\begin{tikzpicture} 
		\node[inner sep=0pt] () at (0,0){\includegraphics[scale=0.5]{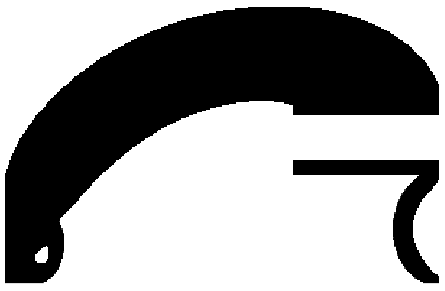} };
		\DrawBoundingBox[gray]
	\end{tikzpicture}
	\caption{Optimized pressure chamber design with the projection filter. $C =61.87$}\label{fig:pressurechamberprojected}
\end{figure}
We also implement the projection filter for the pressure chamber design problem (Sec.~\ref{Sec:Prechamdesign}) After incorporating the aforementioned changes, the 100-line code is called with the projection filter for the pressurized chamber with volume fraction 0.4 as
\begin{lstlisting}[basicstyle=\scriptsize\ttfamily,breaklines=true,numbers=none,frame=tb]
	TOPress(300,200,0.40,3,6,0.1,10,256,1,250)
\end{lstlisting}	
The optimized design with material layout is depicted in Fig.~\ref{fig:pressurechamberprojected}. $M_\text{nd} = 3.1\times\SI{e-2}{}$\% is found. Thus, the optimized pressure chamber design (Fig.~\ref{fig:pressurechamberprojected}) is close to a 0-1 solution. The topology of the optimized design (Fig.~\ref{fig:pressurechamberprojected}) is different than that obtained without the projection filter (Fig.~\ref{fig:pressurechamb_mat_pre_ls_1_volf_40}). The cardinal reason could be that the projection filter changes the search direction, leading to a different optimum point and, thus, a different optimized topology.

\subsection{Other extensions}
The numerical experiments performed in the above subsections suggest that \texttt{TOPress} can be readily modified to determine optimized designs for different pressure loadbearing structures. Problems in Figs.~\ref{fig:2Darcproblem}, \ref{fig:Bridge_structure}, \ref{fig:Arch_ext_structure} and \ref{fig:piston_head_str} can also be solved by exploiting their vertical symmetric nature, we, however, use the full domain to check any deviation from symmetric. Extending the code for three-dimensional problems requires modifications, though the structure of the code may remain the same. The 100-line code can also be extended to design pressure-driven compliant mechanisms per~\cite{kumar2020topology,kumar2021topology}. Further, toward additional physics, one can develop the code for thermal-fluid problems with design-dependent deformation~\citep {zhao2021topology}. Furthermore, one can extend the code with advanced constraints, e.g., buckling, stress, etc., as per the requirement.

\section{Concluding remarks}\label{Sec:Sec4}
This paper presents a compact 100-line MATLAB code, \texttt{TOPress}, for topology optimization of structures involving design-dependent pressure loads using the method of moving asymptotes. Such loads alter their magnitude, direction, and location with topology optimization's progress; dealing with them becomes a challenging and involved task. \texttt{TOPress}, developed based on the method first proposed in~\cite{kumar2020topology}, aims to facilitate the newcomers' and students' learning and provide a potential platform for extending and developing codes/strategies for different applications involving design-dependent pressure loads. The 100-line code and its simple extensions for different problems are explained in detail. A subroutine to plot the pressure field with material layout is also provided. The code is provided in Appendix~\ref{Sec:TOPress}. Extending the code for three-dimensional problems requires modification on multiple fronts; however, the structure of the code may remain the same. The code can also be extended for the design problems with advanced constraints, e.g., buckling, stress, etc. 

In conjunction with the drainage term, the Darcy law is employed to model the pressure load, wherein the flow coefficient is defined using a smooth Heaviside function per~\cite{kumar2020topology}. The consistent nodal loads corresponding to the obtained pressure field are determined. The approach is qualified using three preliminary numerical examples, which indicate that the Darcy law with the drainage term gives the appropriate (realistic) and accurate pressure load modeling and nodal force calculation. The approach facilitates computationally cheap determination of the load sensitivities, which can be switched on/off in the provided code.

Compliance of the structure is minimized with the given volume fraction. Benchmark problems involving pressure load are solved with and without load sensitivities. Including load sensitivities within the formulation can affect the optimized designs' shape and topology, and thus the final objective values. One can obtain relatively better-performing pressure loadbearing structures using the load sensitivities. Based on the numerical examples we perform, it is recommended to consider load sensitivities while optimizing such problems; neglecting them cannot be considered  physically correct and a welcomed idea. \texttt{TOPress} contains six main parts which are described in detail.  We believe  the newcomers, students, and researchers will benefit from the provided codes and the paper, and they will extend/use the codes for their research works involving design-dependent pressure (pneumatic) loads.

\section*{Replication of results}
The 100-line code is explained in detail and provided in Appendix~\ref{Sec:TOPress}. In addition,  TOPress and its extensions are supplied as supplemental material to reproduce the solutions. If anything is missing can be obtained directly from the author.
\section*{Acknowledgment}
The author thanks Prof. Matthijs Langelaar and Prof. Ole Sigmund for discussions on the method in the past and Prof. Krister Svanberg (krille@math.kth.se) for providing MATLAB codes of the MMA optimizer. 
\section*{Compliance to ethical standards}
\subsection*{Conflicts of interest}
The author declares no conflicts of interest.
\subsection*{Ethical approval}
This article does not contain any studies with human participants or animals performed by the author.

\begin{appendices}
	\numberwithin{equation}{section}
	\section{Expressions for flow and transformation matrices: $\mathbf{\mathbf{K}}_\text{p}^i$, $\mathbf{K}^i_\text{Dp}$ and $\mathbf{T}_i$} \label{Sec:APPExpression}
	The expressions for the flow matrices $\mathbf{\mathbf{K}}_\text{p}^i$ and $\mathbf{K}^i_\text{Dp}$, and  transformation matrix $\mathbf{T}_i$ are provided for element~$i$ herein. 
	
\begin{figure}[h!]
		\centering
		\includegraphics[scale=1]{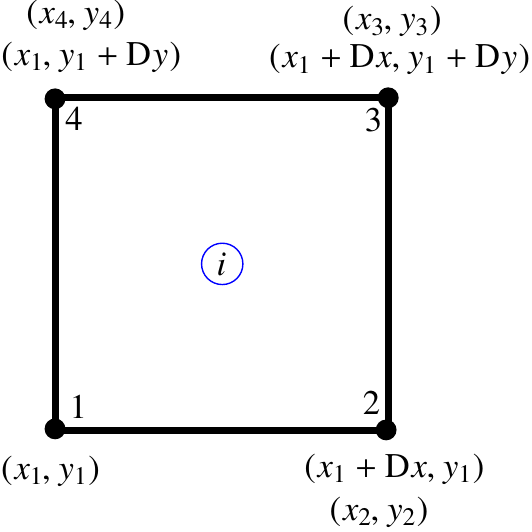}
		\caption{Nodal coordinates of element~$i$}
		\label{fig:AppFig1}
	\end{figure}

	Figure~\ref{fig:AppFig1} displays the local node numbers with nodal coordinates $(x_i,\,y_i)$, $(x_i +\text{D}x,\,y_i)$, $(x_i +\text{D}x,\,y_i+\text{D}y)$, and $(x_i ,\,y_i+\text{D}y)$ for element~$i$ . 
	As mentioned earlier, we employ the bi-linear shape functions for the finite element analysis. To determine,  $\mathbf{\mathbf{K}}_\text{p}^i$ and $\mathbf{K}^i_\text{Dp}$ as written in \eqref{Eq:FEAformulation} and $\mathbf{T}_i$ as provided in \eqref{Eq:Forcepressureconversion}, numerical integration approach using the Gauss points is employed~\citep{zienkiewicz2005finite}. Mathematically,
	\begin{equation}\label{Eq:Appen_expressions}
		\begin{aligned}
			\mathbf{\mathbf{K}}_\text{p}^i &= \int_{\Omega_e} K~ \mathbf{B}^\top_\text{p} \mathbf{B}_\text{p} \text{d} V \\&= t_i\int_{-1}^{1}\int_{-1}^{1} \left( K~ \mathbf{B}^\top_\text{p} \mathbf{B}_\text{p}\right) \det \mathbf{J} d\xi d\zeta \\
			\mathbf{K}^i_\text{Dp} &=  \int_{\Omega_e}D ~\mathbf{N}^\top_\text{p} \mathbf{N}_\text{p} \text{d} V \\&= t_i\int_{-1}^{1}\int_{-1}^{1} \left( D ~\mathbf{N}^\top_\text{p} \mathbf{N}_\text{p}\right) \det \mathbf{J} d\xi d\zeta\\
			\mathbf{T}_i &=-\int_{\mathrm{\Omega}_e} \trr{\mathbf{N}}_\mathbf{u} \mathbf{B}_\text{p}  \text{d} V =t_i\int_{-1}^{1}\int_{-1}^{1} \trr{\mathbf{N}}_\mathbf{u} \mathbf{B}_\text{p}  \det \mathbf{J} d\xi d\zeta
		\end{aligned}
	\end{equation}
	where $\mathbf{J}$ is the Jacobian matrix given as
	\begin{equation}
		\mathbf{J}=	\frac{1}{4}
		\begin{bmatrix}
			J11 &\quad J12\\
			J21 & \quad J22,
		\end{bmatrix}
	\end{equation}
	where $J11 = -(1-\zeta)x_1 + (1-\zeta) x_2 + (1+\zeta)x_3 - (1+\zeta)x_4$, $J12 = \quad -(1-\zeta)y_1 + (1-\zeta) y_2 + (1+\zeta)y_3 - (1+\zeta)y_4$, $J21 = -(1-\xi)x_1 + (1-\xi) x_2 + (1+\xi)x_3 - (1+\xi)x_4,$ and $J22 = -(1-\xi)y_1 + (1-\xi) y_2 + (1+\xi)y_3 - (1+\xi)y_4$,
	with $N_k = \frac{1}{4}(1+\xi\xi_k)(1+ \zeta\zeta_k)$, where $(\xi_k,\,\zeta_k)$ are natural coordinates  with (-1,\,-1), (1,\,-1), (1,\,1) and (-1,\,1) for node $k$. \eqref{Eq:Appen_expressions} yields the following with $2\times2$ Gauss-quadrature rule:
	
	\begin{equation}
		\mathbf{\mathbf{K}}_\text{p}^i=\frac{t_i}{6 Dx Dy}
		\begin{bmatrix}
			K_{p1} &\quad K_{p2} &\quad -0.5K_{p1}&\quad K_{p3}	\\
			&\quad K_{p1} &\quad K_{p3}&\quad -0.5K_{p1}\\
			&\text{Sym.} &K_{p1} &\quad  K_{p2}\\& & &\quad K_{p1}
		\end{bmatrix}
	\end{equation}
	where $K_{p1} = 2(Dx^2+ Dy^2)$,\, $K_{p2} = Dx^2 -2Dy^2$,\,$K_{p3} = Dy^2 - 2Dx^2$. 
	
	\begin{equation}
		\mathbf{\mathbf{K}}_\text{Dp}^i=\frac{t_i Dx Dy}{36}
		\begin{bmatrix}
			4 &\quad 2 &\quad 1 &\quad2	\\
			&\quad 4 &\quad 2 &\quad 1\\
			& \text{Sym.} &\quad4 &\quad2\\
			
			& & &\quad 4
		\end{bmatrix}
	\end{equation}
	and
	\begin{equation}
		\mathbf{T}_i=\frac{t_i}{12}
		\begin{bmatrix}
			-2Dy &\quad 2Dy &\quad Dy &\quad -Dy	\\
			-2Dx  &\quad -Dx &\quad Dx &\quad 2Dx\\
			-2Dy &\quad 2Dy &\quad Dy &\quad -Dy\\
			-Dx&\quad -2Dx &\quad 2Dx &\quad Dx\\
			-Dy	&\quad Dy &\quad 2Dy &\quad -2Dy\\
			-Dx&\quad -2Dx &\quad 2Dx &\quad Dx\\
			-Dy	&\quad Dy &\quad 2Dy &\quad -2Dy\\
			-2Dx  &\quad -Dx &\quad Dx &\quad 2Dx
		\end{bmatrix}
	\end{equation}
	For the square elements which we consider to parameterize the design domain with $t_i =1$ (plane-stress cases). The expressions for $\mathbf{\mathbf{K}}_\text{p}^i$, $\mathbf{\mathbf{K}}_\text{Dp}^i$ and 	$\mathbf{T}_i$ transpire to
	\begin{equation}
		\mathbf{\mathbf{K}}_\text{p}^i=\frac{1}{6}
		\begin{bmatrix}
			4 &\quad -1 &\quad -2&\quad -1	\\
			&4 &\quad -1&\quad -2\\
			&\text{Sym.} &4 &\quad -1\\& & &4
		\end{bmatrix}
	\end{equation}
	\begin{equation}
		\mathbf{\mathbf{K}}_\text{Dp}^i=\frac{1}{36}
		\begin{bmatrix}
			4 &\quad 2 &\quad 1 &\quad2	\\
			&\quad 4 &\quad 2 &\quad 1\\
			& \text{Sym.} &\quad4 &\quad2\\
			
			& & &\quad 4
		\end{bmatrix}
	\end{equation}
	and
	\begin{equation}
		\mathbf{T}_i=\frac{1}{12}
		\begin{bmatrix}
			-2 &\quad 2 &\quad 1 &\quad -1	\\
			-2  &\quad -1 &\quad 1 &\quad 2\\
			-2 &\quad 2 &\quad 1 &\quad -1\\
			-1&\quad -2 &\quad 2 &\quad 1\\
			-1	&\quad 1 &\quad 2 &\quad -2\\
			-1&\quad -2 &\quad 2 &\quad 1\\
			-1	&\quad 1 &\quad 2 &\quad -2\\
			-2  &\quad -1 &\quad 1 &\quad 2
		\end{bmatrix}
	\end{equation}
	\onecolumn
	\section{The MATLAB code: \texttt{TOPress}}\label{Sec:TOPress}
	\begin{lstlisting}[basicstyle=\scriptsize\ttfamily,breaklines=true]
function TOPress(nelx,nely,volfrac,penal,rmin,etaf,betaf,lst,maxit)
%% ___PART 1.____________________________MATERIAL AND FLOW PARAMETERS
E1 = 1;
Emin = E1*1e-6;
nu = 0.30;
[Kv,epsf,r,Dels] = deal(1,1e-7,0.1,2);                  % flow parameters
[Ds, kvs]= deal((log(r)/Dels)^2*epsf,Kv*(1 - epsf));    % flow parameters
%% ____PART 2._______________FINITE ELEMENT ANALYSIS PREPARATION and NON-DESIGN DOMAIN
[nel,nno] = deal(nelx*nely, (nelx+1)*(nely+1));
nodenrs = reshape(1:(1+nelx)*(1+nely),1+nely,1+nelx);
edofVec = reshape(2*nodenrs(1:end-1,1:end-1)+1,nelx*nely,1);
Udofs = repmat(edofVec,1,8)+repmat([0 1 2*nely+[2 3 0 1] -2 -1],nelx*nely,1);
[Lnode,Rnode,]= deal(1:nely+1, (nno-nely):nno);
[Bnode,Tnode]= deal((nely+1):(nely+1):nno, 1:(nely+1):(nno-nely));
[Pdofs,allPdofs, allUdofs] = deal(Udofs(:,2:2:end)/2,1:nno,1:2*nno);
iP = reshape(kron(Pdofs,ones(4,1))',16*nel,1);
jP = reshape(kron(Pdofs,ones(1,4))',16*nel,1);
iT = reshape(kron(Udofs,ones(4,1))',32*nel,1);
jT = reshape(kron(Pdofs,ones(1,8))',32*nel,1);
iK = reshape(kron(Udofs,ones(8,1))',64*nel,1);
jK = reshape(kron(Udofs,ones(1,8))',64*nel,1);
Kp = 1/6*[4 -1 -2 -1;-1 4 -1 -2; -2 -1 4 -1; -1 -2 -1 4]; % flow matrix: Darcy Law
KDp = 1/36*[4 2 1 2; 2 4 2 1; 1 2 4 2; 2 1 2 4];          % Drainage matrix
Te = 1/12*[-2 2 1 -1;-2 -1 1 2;-2 2 1 -1;-1 -2 2 1;-1 1 2 -2; -1 -2 2 1; -1 1 2 -2; -2 -1 1 2]; % tran matrix
A11 = [12  3 -6 -3;  3 12  3  0; -6  3 12 -3; -3  0 -3 12];
A12 = [-6 -3  0  3; -3 -6 -3 -6;  0 -3 -6  3;  3 -6  3 -6];
B11 = [-4  3 -2  9;  3 -4 -9  4; -2 -9 -4 -3;  9  4 -3 -4];
B12 = [ 2 -3  4 -9; -3  2  9 -2;  4  9  2  3; -9 -2  3  2];
ke = 1/(1-nu^2)/24*([A11 A12;A12' A11]+nu*[B11 B12;B12' B11]);             %stiffness matrix
IFprj=@(xv,etaf,betaf)((tanh(betaf*etaf) + tanh(betaf*(xv-etaf)))/...      %projection function
		(tanh(betaf*etaf) + tanh(betaf*(1 - etaf))));
dIFprj=@(xv,etaf,betaf) betaf*(1-tanh(betaf*(xv-etaf)).^2)...
		/(tanh(betaf*etaf)+tanh(betaf*(1-etaf)));                    % derivative of the projection function
[NDS, NDV ] = deal( [], [] );
act = setdiff((1 : nel)', union( NDS, NDV ));
%% ____PART 3.______PRESSURE & STRUCTURE B.C's, LOADs
[PF, Pin] =deal(0.00001*ones(nno,1),1);        %pressure-field preparation
PF([Tnode,Lnode,Rnode]) = 0; PF(Bnode) = Pin;  % applying pressure load
fixedPdofs = allPdofs(PF~=0.00001);            % given P-dofs
freePdofs  = setdiff(allPdofs,fixedPdofs);     % free P-dofs
pfixeddofsv = [fixedPdofs' PF(fixedPdofs)];    % p-fixed and its value
fixedUdofs = [2*Bnode(1)-1  2*Bnode(1)  2*Bnode(end)-1 2*Bnode(end)]; %fixed displ.
freeUdofs = setdiff(allUdofs,fixedUdofs);      % free dofs for displ.
[U, lam1] = deal(zeros(2*nno,1),zeros(nno,1)); %lam1:Lagrange mult.
%% ___PART 4._________________________________________FILTER PREPARATION
[dy,dx] = meshgrid(-ceil(rmin)+1:ceil(rmin)-1,-ceil(rmin)+1:ceil(rmin)-1);
h = max( 0, rmin-sqrt( dx.^2 + dy.^2 ) );      % conv. kernel
Hs = imfilter(ones(nely, nelx ), h);           % matrix of weights (filter)
%% ___PART 5.__________________________MMA OPTIMIZATION PREPARATION & INITIALIZATION
[x,dVol0] = deal(zeros(nel,1),ones(nel,1)/(nel*volfrac)); %design var. and vol. cont. der.
x(act) = (volfrac*(nel-length(NDV))-length(NDS) )/length(act); x(NDS) = 1; % updating
[nMMA,mMMA,xphys,xMMA,mvLt] = deal(length(act),1,x,x(act),0.1); % diff. varbls.
[xminvec,xmaxvec] = deal(zeros(nMMA,1),ones(nMMA,1));           %Min. & Max
[low, upp] = deal(xminvec,xmaxvec);                             % Low and Upp limits MMA
[cMMA,dMMA, a0, aMMA] = deal(1000*ones(mMMA,1),zeros(mMMA,1),1,zeros(mMMA,1));
[xold1,xold2] = deal(xMMA);
[loop, change] =deal(0,1);                % loop counter and convergence change criteria
dVol = imfilter(reshape(dVol0, nely, nelx)./Hs,h); %filtered volume sensitivity
%% ____PART 6._____________________________________MMA OPTIMIZATION LOOP
while(loop<maxit && change>0.01)
loop = loop + 1;  % Updating the opt. iteration
%___PART 6.1__________SOLVING FLOW BALANCE EQUATION
Kc = Kv*(1-(1-epsf)*IFprj(xphys,etaf,betaf));       %Flow coefficient
Dc = Ds*IFprj(xphys,etaf,betaf);                    %Drainage coefficient
Ae = reshape(Kp(:)*Kc' + KDp(:)*Dc',16*nel,1);      %Elemental flow matrix
AG = (sparse(iP,jP,Ae)+ sparse(iP,jP,Ae)')/2;       %Global flow matrix
Aff = AG(freePdofs,freePdofs);                      %AG for free pressure dofs
PF(freePdofs) = decomposition(Aff,'ldl')\(-AG(freePdofs,fixedPdofs)*pfixeddofsv(:,2));
PF(pfixeddofsv(:,1)) = pfixeddofsv(:,2);            % Final P-field
%__PART 6.2_DETERMINING CONSISTENT NODAL LOADS and GLOBAL Disp. Vector
Ts = reshape(Te(:)*ones(1,nel), 32*nel, 1);         %Elemental transformation matrix
TG = sparse(iT, jT, Ts);                            %Global transformation matrix
F = -TG*PF;                                         % Dertmining nodal forces
E = Emin + xphys.^penal*(E1 - Emin);                %Material interpolation
Ks = reshape(ke(:)*E',64*nel,1);                    %Elemental stiffness matrix
KG = (sparse(iK,jK,Ks) + sparse(iK,jK,Ks)')/2;      %Global stiffnes matrix
U(freeUdofs) = decomposition(KG(freeUdofs,freeUdofs),'chol','lower')\F(freeUdofs); %Global Disp. Vect.
%__PART 6.3__OBJECTIVE, CONSTRAINT and THEIR SENSITIVITIES COMPUTATION
obj = U'*KG*U;                                      % determining objective
lam1(freePdofs) = (2*U(freeUdofs)'*TG(freeUdofs,freePdofs))/Aff; % Lagrange mult.
objsT1 = -(E1 - Emin)*penal*xphys.^(penal - 1).*sum(([U(Udofs)]*ke).*[U(Udofs)],2);
dC1k = -dIFprj(xphys,etaf,betaf).* sum((lam1(Pdofs)*(kvs*Kp)) .* PF(Pdofs),2);
dC1h =  dIFprj(xphys,etaf,betaf).* sum((lam1(Pdofs)*(Ds*KDp)) .* PF(Pdofs),2);
objsT2 = dC1k + dC1h; objsens = (objsT1 + lst*objsT2); % final sensitivities
Vol = sum(xphys)/(nel*volfrac)-1;                      % volume fraction
if(loop ==1), normf = 1000/(obj);save normf normf;end, load normf normf;
objsens = imfilter(reshape(objsens*normf, nely, nelx)./Hs,h); % Obj. sens.
%___PART 6.4______________________SETTING and CALLING MMA OPTIMIZATION
xval = xMMA;
[xminvec, xmaxvec]= deal(max(0, xval - mvLt),min(1, xval + mvLt));
[xmma,~,~,~,~,~,~,~,~,low,upp] = mmasub(mMMA,nMMA,loop,xval,xminvec,xmaxvec,xold1,xold2, ...
		obj*normf,objsens(act)',objsens(act)'*0,Vol,dVol(act)',dVol(act)'*0,low,upp,a0,aMMA,cMMA,dMMA); % call MMA
[xold2,xold1, xnew]= deal(xold1, xval,xmma);    % updating
change = max(abs(xnew-xMMA));                   % Calculating change
xMMA = xnew;   xphys(act) = xnew;xphys = imfilter(reshape(xphys, nely, nelx),h)./Hs;
xphys = xphys(:); xphys(NDS)=1; xphys(NDV)=0;   % updating xphys
%___PART 6.5_____________________________Printing and plotting results
fprintf(' It.:%5i Obj.:%11.4f Vol.:%7.3f ch.:%7.3f\n',loop,obj*normf,mean(xphys),change);
colormap(gray); imagesc(1-reshape(xphys, nely, nelx));caxis([0 1]);axis equal off;drawnow;
end
	\end{lstlisting}
	\lstdefinestyle{nonumbers}
	{numbers=none}
	\begin{lstlisting}[style=nonumbers]
%%%%%%%%%%%%%%%%%%%%%%%%%%%%%%%%%%%%%%%%%%%%%%%%%%%%%%%%%%%%%%%%%%%%%%%%%%%%%%%%%%%%%%
%    The code, TOPress,  is provided for  pedagogical purposes. A  detailed          %
%    description is presented in the paper:"TOPress: a MATLAB implementation for     %
%    topology optimization of structures subjected to design-dependent pressure      %
%    loads"   Structural and Mutltidisciplinary Optimization, 2023.                  %
%                                                                                    %
%    One can download the code and its extensions for the different problems         %
%    from the online supplementary material and also from:                           %
%                                      https://github.com/PrabhatIn/TOPress          %
%                                                                                    %
%    Please send your comment to: pkumar@mae.iith.ac.in                              %
%                                                                                    %
%    One may also refer to the following two papers for more detail:                 % 
%                                                                                    %
%    1. Kumar P, Frouws JS, Langelaar M (2020) Topology optimization of fluidic      %
%    pressure-loaded structures and compliant mechanisms using the Darcy method.     %
%    Structural and Multidisciplinary Optimization 61(4):1637-1655                   %
%    2. Kumar P, Langelaar M (2021) On topology optimization of design-dependent     % 
%    pressure-loaded three-dimensional structures and compliant mechanisms.          %
%    International Journal for Numerical Methods in Engineering 122(9):2205-2220     %
%                                                                                    %
%    Disclaimer:                                                                     %
%    The author does not guarantee that the code is free from erros but reserves     %
%    all rights. Further, the author shall not be liable in any event caused by      % 
%    use of the above 100-line code and its extensions                               %
%                                                                                    %
%%%%%%%%%%%%%%%%%%%%%%%%%%%%%%%%%%%%%%%%%%%%%%%%%%%%%%%%%%%%%%%%%%%%%%%%%%%%%%%%%%%%%%
	\end{lstlisting}
\end{appendices}

\end{document}